\documentclass[useAMS,usenatbib]{mn2e}

\title[]{A massive bubble of extremely metal poor gas around a collapsing Ly$\alpha$ blob at z=2.54}
\author[Andrew Humphrey et al.]{A. Humphrey$^{1,2}$, L. Binette$^{3,4}$, M. Villar-Mart\'{i}n$^{5}$, I. Aretxaga$^{2}$, P. Papaderos$^{1}$\\
$^{1}$Centro de Astrof\'{i}sica da Universidade do Porto, Rua das Estrelas, 4150-762, Porto, Portugal\\
$^{2}$Instituto Nacional de Astrof\'{i}sica, \'Optica y Electr\'onica, Luis Enrique Erro 1, Sta. Ma. Tonantzintla, Puebla, M\'{e}xico\\
$^{3}$Instituto de Astronom\'{i}a, Universidad Nacional Aut\'onomo de M\'exico, Ap. 70-264, 04510 M\'exico, D.F., M\'exico\\
$^{4}$D\'{e}partement de Physique, de G\'{e}nie Physique et d'Optique, Universit\'{e} Laval, Qu\'{e}bec, QC, G1V\,0A6, Canada\\
$^{5}$Centro de Astrobiolog\'{i}a (INTA-CSIC), Carretera de Ajalvir, km 4, 28850 Torrej\'on de Ardoz, Madrid, Spain
}

\begin{document}

\date{Accepted 20 Sept 2012.
      Received 20 Sept 2012;
      in original form 9 Aug 2012.}

\pagerange{\pageref{firstpage}--\pageref{lastpage}}
\pubyear{2012}

\maketitle

\label{firstpage}

\begin{abstract}
Using long-slit optical spectroscopy obtained at the 10.4 m Gran Telescopio Canarias, we have examined the gaseous environment of the radio-loud quasar TXS 1436+157 (z=2.54), previously known to be associated with a large Ly$\alpha$ nebula and a spatially extended Ly$\alpha$-absorbing structure.  From the Ly$\alpha$ nebula we measure kinematic properties consistent with infall at a rate of $\sim$10-100 $M_{\odot}$ yr$^{-1}$ -- more than sufficient to power a quasar at the top of the luminosity function.  

The absorbing structure lies outside of the Ly$\alpha$ nebula, at a radius of $\ga$40 kpc from the quasar.  Against the bright unresolved continuum and line emission from the quasar, we detect in absorption the NV $\lambda\lambda$1239,1241, CIV$\lambda\lambda$1548,1551 and SiIV $\lambda\lambda$1394,1403 doublets, with no unambiguous detection of absorption lines from any low-ionization species of metal.  The metal column densities, taken together with the HI column density measurement from the literature, indicate that the absorbing gas is predominantly ionized by the quasar, has a mass of hydrogen of $\ga 1.6\times10^{11}$ M$_{\odot}$, a gas density of $\le$18 cm$^{-3}$, a line of sight thickness of $\ge$18 pc, and a covering factor approaching unity.  While this absorbing structure is clearly not composed of pristine gas, it has an extremely low metallicity, with ionization models providing a 3$\sigma$ limit of 12+log(O/H)$\le$7.3.  

To explain these results, we discuss a scenario involving starburst-driven superbubbles and the creation of infalling filaments of cold gas which fuel/trigger the quasar.  We also discuss the possibility of detecting large-scale absorbers such as this in emission when illuminated by a powerful quasar.  
\end{abstract}


\section{Introduction}
Ly$\alpha$ nebulae at high redshift (z$\ga$2:
  HzLAN), sometimes also known as Ly$\alpha$ `blobs', `halos' or `fuzz', are
  prodigious sources of HI Ly$\alpha$ photons, with luminosities up to
  $\sim$10$^{45}$ erg s$^{-1}$, and sizes often
  exceeding $\sim$100 kpc (e.g. Steidel et al. 2000).  HzLANs are
not associated with a common type of galaxy: they are variously associated with 
UV-selected star forming galaxies, dusty star-forming galaxies, powerful active 
galactic nuclei, or with no apparent optical galaxy at all (e.g. McCarthy et al. 1987; Reuland et al. 2003; 
Webb et al. 2009; Steidel et al. 2011; Bridge et al. 2012).  A
  complete understanding of the nature of HzLANs has not yet been
  reached, with key properties such as their dominant power source(s)
  and their space density remaining particularly poorly
  understood. Nevertheless, HzLANs are thought to
  represent an important phase of mass assembly, and understanding this energetic phenomenon promises 
to yield significant insights into the physics of massive 
galaxy formation (e.g., Mori \& Umemura et al. 2006).  

Associated absorbers offer a complementary means by which the gaseous environment of galaxies can be studied (e.g. Pettini et al. 2002; Fosbury et al. 2003; Wilman et al. 2005).  A substantial fraction of radio-loud, powerful active galaxies at high-z show spatially extended, narrow (FWHM$\la$500 km s$^{-1}$) absorption lines, usually HI Ly$\alpha$, close to the redshift of the galaxy; the observational characteristics of the absorbers suggest they are giant shells of gas, with radii $\ga$50 kpc, and are distinct structures from the narrow line emitting nebulae that are also usually associated with powerful active galaxies (R\"ottgering et al. 1995a; van Ojik et al. 1997; Binette et al. 2000; Jarvis et al. 2003; Wilman et al. 2004; Humphrey et al. 2008a).  Intriguingly, in the case of high-z radio galaxies there is a strong anticorrelation between the projected size of the radio source and the detection of associated HI absorbers, with $\sim$90 per cent of high-z radio galaxies that have small radio sources ($<$50 kpc) showing an associated absorber, falling to $\sim$25 percent for large radio sources ($>$50 kpc: van Ojik et al. 1997).  This result has been interpreted in terms of (a) creation of an expanding shell of gas swept up from the interstellar medium of the host galaxy by the expanding radio source (e.g. Krause 2002), or (b) pre-existing gaseous shells produced by powerful starbursts, which are over-taken and disrupted by the expanding radio source (e.g. Jarvis et al. 2003).  It is not yet clear whether the absorbing shells are in infall or outflow (e.g. Jarvis et al. 2003; Humphrey et al. 2008a).  At least a few of the HI absorbers also contain carbon, as shown by their detection in CIV $\lambda$1549 absorption lines (e.g. Binette et al. 2000; Jarvis et al. 2003; Humphrey et al. 2008a).  Binette et al. (2000, 2006) and Jarvis et al. (2003) have used ionization models to place important constraints on the metallicity, the source of ionization and the density of the HI and CIV absorbers detected in front of radio galaxies USS 0200+015 (z=2.23) and USS 0943-242 (z=2.92).  Interestingly, Vernet et al. (2001) detected various narrow absorption lines from low-ionization or neutral species of C, O, S and Si against the UV continuum emission of several high-z radio galaxies, and noted a tendency for the absorption lines to be stronger in the galaxies with more highly polarized UV continuum emission.  The possible relationship between these low ionization metal absorption lines and the spatially extended HI absorbers is not yet clear.  

Quasars provide a useful alternative perspective from which to examine narrow associated absorbers.  The relatively unobscured view to their highly luminous central engines significantly enhance the detectability of narrow absorption lines that would, as in type II QSOs or radio galaxies, otherwise lie in spectral regions with no strong continuum or line emission.  The absorbers also ought to have a relatively unobscured view of the AGN, raising the possibility that some of the narrow absorbers associated with quasars may be photoionized by the hard radiation field of the AGN; this may result in the detection of absorption lines from more numerous metal species than neutral or stellar-photoionized absorbers, greatly aiding analyses of excitation and metallicity.  Moreover, by examining the properties of associated absorbers as a function of AGN and jet orientation, information about the geometry of the absorbers can be obtained.  Indeed, Baker et al. (2002), in their study of associated absorption in radio-loud quasars, found that the detection of associated narrow CIV absorbers is orientation-dependent, being detected more frequently in the lobe-dominated or steep spectrum quasars than in the flat-spectrum or core-dominated quasars, suggesting that the absorbing gas lies away from the axis of the radio jet.  The authors also noted an anticorrelation between the equivalent width of narrow CIV absorption lines and the projected diameter of the radio source, broadly similar to the trend identified by van Ojik et al. (1997) for high-z radio galaxies.  

We have begun a programme of spectroscopic observations, designed to help elucidate the nature of Ly$\alpha$ nebulae and HI absorbers associated with powerful active galaxies at high-z.  By targeting active galaxies known to show extended Ly$\alpha$ nebulae, extended HI absorption close to the redshift of the galaxy, and a very bright unresolved emission component ($\ga$10$^{-17}$ erg s cm$^{-2}$ \AA$^{-1}$), we aim to obtain the geometric, ionization and chemical properties of individual absorbers.  This approach also mitigates some problems inherent to previous studies, such as the lack of sufficiently bright background emission to readily detect absorption lines other than HI Ly$\alpha$ or CIV, or the lack of an extended background source to provide information about the size and location of the absorber.  Based on this idea, Humphrey (2008a) employed deep longslit spectroscopy from Keck II, together with integral field spectroscopy from the Very Large Telescope, to examine the properties of a spatially extended HI absorber associated with the reddened radio-loud quasar MRC 2025-218.  The authors traced the HI absorption feature across a projected area of $\sim$30 kpc $\times$ 40 kpc, with very little variation in kinematic properties therein, which implies the absorber is a coherent structure located at least 20 kpc from the AGN.  Against the unresolved continuum and line emission from the active nucleus, narrow absorption lines of CI, CII, CIV, NV, OI, SiII, SiIV, AlII and AlIII were detected at the redshift of the HI absorber.  The relative column densities of the metal absorption lines indicate the absorbing gas is photoionized by the active nucleus, has a gas density of $\ga$10 cm$^{-3}$, and is enriched in metals.  

Following on from Humphrey et al. (2008a), in this paper we present long-slit spectroscopic observations obtained at the 10.4 m Gran Telescopio Canarias of the $R=19.6$ quasar TXS 1436+157 at z=2.537.  With an 4.7 GHz flux density of 47 mJy, this quasar qualifies as radio-loud, and its radio source shows a steep spectrum and a lobe-dominated morphology (R\"ottgering 1993; Carilli et al. 1997).  Of particular relevance to this programme, TXS 1436+157 is also known to be embedded within a $\sim$90 kpc diameter nebula with a Ly$\alpha$ luminosity of $\ge$ 2.2$\times$10$^{44}$ erg s$^{-1}$ (van Ojik et al. 1997: R\"ottgering et al. 1997), and shows a strong and spatially extended Ly$\alpha$ absorption feature within its Ly$\alpha$ emission line profile (van Ojik et al. 1997).  The primary aim of this paper is to elucidate the properties of the Ly$\alpha$ nebula and the associated absorber.  We assume $H_{0}$=71 km s$^{-1}$ Mpc$^{-1}$, $\Omega_{\Lambda}$=0.73 and $\Omega_{m}$=0.27.  At the redshift of TXS 1436+157, 1\arcsec corresponds to 8.2 kpc.  

\begin{figure}
\includegraphics{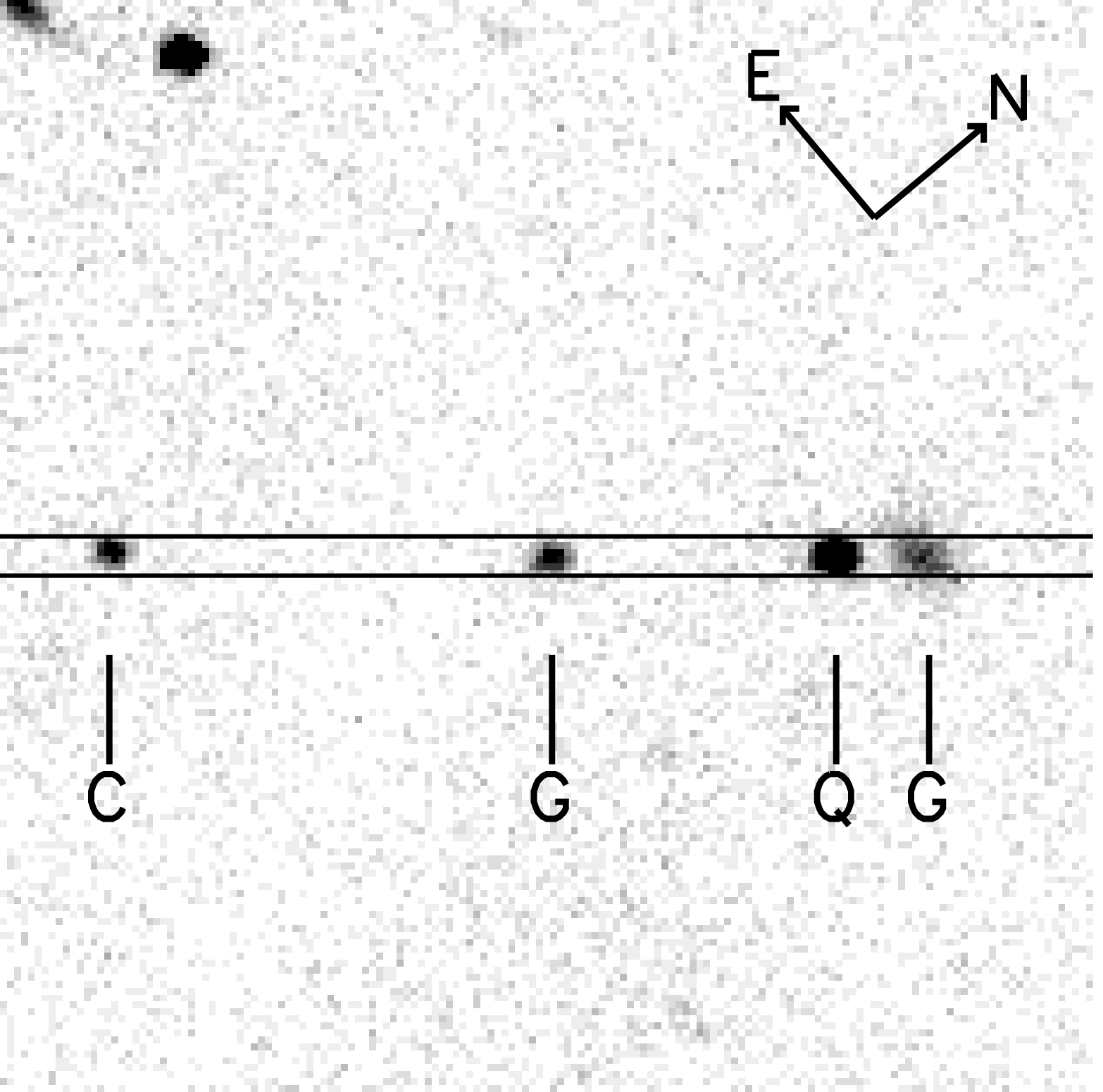}
\includegraphics{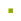}
\vspace{3.35in}
\caption{Image of the 40\arcsec $\times$ 40\arcsec field around TXS 1436+157, taken through a Sloan $r$ filter with an exposure time of 20 s.  The position of the long slit is shown.  Four discrete sources were within the slit: the z=2.54 quasar TXS 1436+157 (Q); two galaxies which we know now to be at z=0.160 and z=0.159, respectively (G); and a continuum only source of unknown redshift (C).}
\label{fig1}
\end{figure}

\section{Observations and data reduction}

Long-slit spectroscopic observations of TXS 1436+157 were obtained in service mode on 2010 May 16 and May 21, using the Optical System for Imaging and low Resolution Integrated Spectroscopy (OSIRIS) instrument at 10.4 m Gran Telescopio Canarias.  The observing conditions were dark and photometric.  The total integration of 7590 s was split into 3 integrations of 2530 s each, one of which was carried out on 2010 May 16, while the other two were carried out on 2010 May 21.  Although the FWHM of the seeing disc was 1\arcsec during all of the observations, the primary mirror segments were not optimally aligned during the observations, which resulted in a poor spatial point-spread function {\it along the slit} (FWHM=2.3\arcsec, 1.6\arcsec and 2.3\arcsec).  The R1000B grism was used, yielding a spectral range of 3700-7000 \AA.  A 1.23\arcsec wide slit was used, and was oriented at a position angle of 50$^{\circ}$ North through East, so as to include TXS 1436+157 and three additional sources that are within $\sim$30\arcsec of the quasar, and which are closely aligned with the major axis of the quasar's radio emission (R\"ottgering et al. 1995).  Fig. 1 shows our 20 s acquisition image taken through a Sloan $r$ filter, with the slit position overlaid.  The instrument was read-out with 2$\times$2 binning, resulting in a spatial scale of 0.25\arcsec pixel$^{-1}$, and a wavelength scale of 2.22 \AA~pixel$^{-1}$  The data were reduced in the standard way using long-slit reduction software from the IRAF software suite.  Flux calibration was performed using observations of GRW+708247.  

Since the slit width (1.23\arcsec) was significantly larger than the seeing disc FWHM ($\sim$1\arcsec), the instrumental profile (IP) differs between unresolved and extended sources.  For extended sources the FWHM of the IP is 11.0 \AA, measured from night sky lines in the spectra; this IP will be adopted when calculating the FWHM of emission lines from spatially extended line-emitting gas.  For unresolved sources, it is expected to be $\sim$8 \AA, in good agreement with narrow emission lines in the spectra of the intervening star forming galaxies and from the strong absorption lines in the spectrum of the quasar; this IP will be used for absorption lines detected against the unresolved line and continuum emission of the quasar.  

\begin{figure}
\includegraphics{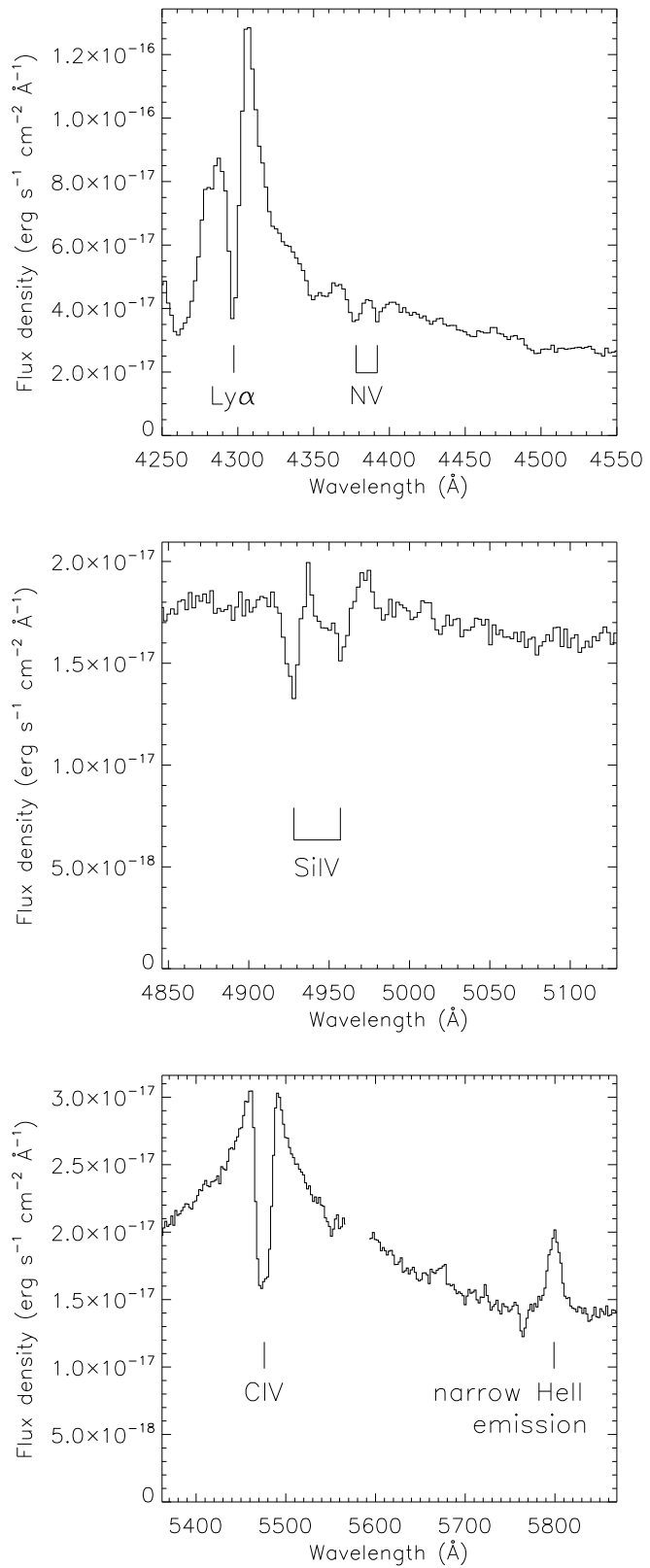}
\vspace{8.55in}
\caption{Sections of the one-dimensional spectrum showing the absorption lines detected in front of TXS 1436+157.  In the lower panel, the gap in the data between CIV and HeII hides a substantial sky-subtraction residual.}
\label{spectra}
\end{figure}

\begin{figure}
\includegraphics{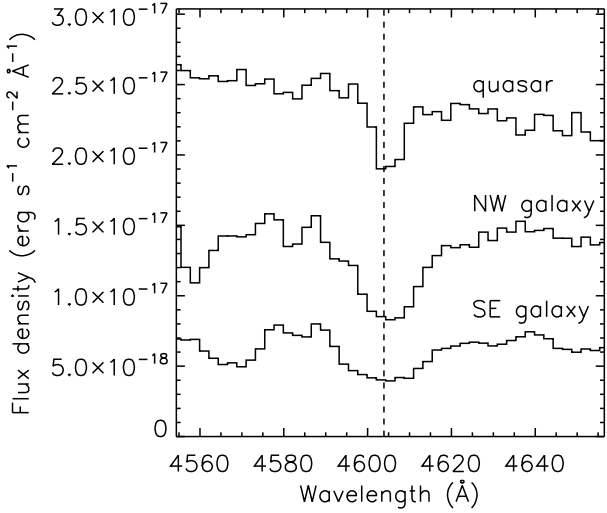}
\vspace{2.85in}
\caption{Sections of the one-dimensional spectra of TXS 1436+157 and the two intervening HII galaxies, with the expected wavelength of the OI $\lambda$1302.2 absorption line marked with a dashed line.  The flux scale is for TXS 1436+157; the spectra of the intervening galaxies have been scaled up and smoothed.  Note that all three of the galaxies show a strong absorption feature close to this wavelength.}
\label{OI_spectra}
\end{figure}

\section{Interloping sources}
The broad band image of TXS 1436+157 presented by R\"ottgering et al. (1995b) shows four tightly aligned sources, including the active galaxy, running along a position angle of $\sim$40$^{\circ}$ (see our Fig. 1).  The radio source shows a roughly similar position angle (Carilli et al. 1997).  Our long slit was positioned so as to encompass all four of these sources, in order to investigate the nature of these aligned optical sources.  Our long slit spectrum reveals that the two closest sources on the sky to TXS 1436+157 are in fact emission line galaxies at intermediate redshifts.  At a distance on the sky of 3\arcsec~ NW, SDSS J143904.77+153121.5 has z=0.159, while SDSS J143905.47+153112.6 at a distance of 11\arcsec~ SE has z=0.160.  Thus, the alignment of these nearby (in projection) sources with the radio source is merely coincidental.  In the case of the source 25\arcsec~ to the SE, we do not detect any strong absorption or emission features, and thus we cannot ascertain its redshift.

\begin{table}
\centering
\caption{Measurements and limits on the absorption lines.  Columns: (1) species; (2) rest-frame wavelength (\AA); (3) velocity shift (km s$^{-1}$) relative to the central wavelength of the nuclear narrow HeII $\lambda$1640 emission line; negative values indicate a blueward shift; the 1$\sigma$ uncertainties are 50-70 km s$^{-1}$; (4) observer-frame equivalent width (\AA); we show for completeness the lower limits for HI Ly$\alpha$ and CIV; (5) oscillator strength of the transition(s); (6) column density of the atom/ion.  The rows beginning 'NV mean' and 'SiIV mean' give 1/$\sigma^2$ weighted means for $N_{N+4}$ and $N_{Si+3}$, respectively. `SiII total' gives a combined upper limit for the 1260.4 \AA~ line and the 1264.7 \AA~ excited fine structure line.  $^{\dag}$The HI velocity and column density is taken from van Ojik et al. (1997).} 
\begin{tabular}{llllll}
\hline
Species & $\lambda_0$& $\Delta$v   & W$_{\lambda}$ & f & $N$ \\  
           & \AA  & km s$^{-1}$  & \AA     & & 10$^{14}$ cm$^{-2}$ \\
(1)        & (2)    & (3)  & (4)          & (6)   & (7) \\
\hline

HI         & 1215.7 & -70$^{\dag}$  &     &  & $\sim$4$\times$10$^{5}$$^{\dag}$  \\
CI,CI*      & 1561.1 &      &        $\le$0.2    & 0.080 & $\le$0.33    \\
CI,CI*      & 1657.2 &      &        $\le$0.4    & 0.14  & $\le$0.33    \\
CII,CII*   & 1335.3 &      &      $\le$0.7    & 0.13  & $\le$1.0    \\
CIV        & 1549.5 & -220 &  $\ge$8.4   & 0.29  & $\ge$3.8      \\
NV         & 1238.8 & -190 & 1.9$\pm$0.2 & 0.16  & 2.47$\pm$0.26 \\
NV         & 1242.8 & -120 &  0.9$\pm$0.2 & 0.078 & 2.39$\pm$0.53 \\
NV mean &          &          &                         &           & 2.45$\pm$0.23 \\
OI         & 1302.2 &         &        $\le$1.2    & 0.049 & $\le$4.6 \\
MgI      & 2026.5 &          & $\le$0.6 &        0.11 & $\le$0.42 \\
SiI &        1425.0 &          & $\le$0.3 &        0.19 &   $\le$0.25 \\
SiI &        1845.5  &         & $\le$0.3 &       0.23 &    $\le$0.12 \\
SiII       & 1260.4 &            &  $\le$0.9    & 1.18  & $\le$0.15    \\
SiII*      & 1264.7 &      &      $\le$0.6    & 1.18  & $\le$0.10    \\
SiII total &           &           & $\le$1.5    &          & $\le$0.25  \\
SiIV       & 1393.8 & -40  &  3.1$\pm$0.3 & 0.52  & 0.98$\pm$0.09 \\
SiIV       & 1402.8 & -170 & 1.9$\pm$0.7 & 0.26  & 1.19$\pm$0.44 \\
SiIV mean &         &          &                       &           & 0.99$\pm$0.09 \\
SI         & 1807.3 &           & $\le$0.6 &     0.11 & $\le$0.52 \\
SII        & 1259.5 &           & $\le$0.6 &     0.016 & $\le$7.5 \\
\hline
\end{tabular}
\label{columns}
\end{table}

\section{Results and Discussion}

\subsection{Measuring $W_{\lambda}$ for absorption lines}

For a formal detection we require (i) that the FWHM of the candidate absorption feature is not smaller than that of the instrumental profile, (ii) that difference between the continuum level and the minimum of the feature is at least twice the local 1$\sigma$ pixel to pixel uncertainty in flux density, and (iii) in the case of doublets, that the wavelength separation and $W_{\lambda}$ are in agreement with their theoretical values.  Since they are doublets and satisfy criterion (iii), our identifications of the NV, SiIV and CIV lines are rather secure ($>$10$\sigma$).  Due to the low spectral resolution of the data, absorption lines were measured using single negative Gaussian profiles, in order to obtain their equivalent width ($W_{\lambda}$).  Uncertainties are derived from uncertainty in the level of the underlying continuum and broad line emission.  

The CIV absorption doublet has altered the overall velocity profile of the CIV line to such an extent that the velocity profile of the unabsorbed emission, and hence $W_{\lambda}$, cannot reliably be reconstruced -- at least at this spectral resolution (Fig. \ref{spectra}).  Our determine a lower limit on $W_{\lambda}$ was measured assuming the two emission peaks of the CIV profile represent the maximum height of the intrinsic (unabsorbed) line profile.  The velocity profile of HI Ly$\alpha$ presents a similar challenge at our spectral resolution.  In this case, we refer to the moderate resolution spectrum and fitting presented by van Ojik et al. (1997), who favoured $N_{HI}\sim$4.0$\times 10^{19}$ cm$^{-2}$.  

\begin{figure*}
\includegraphics{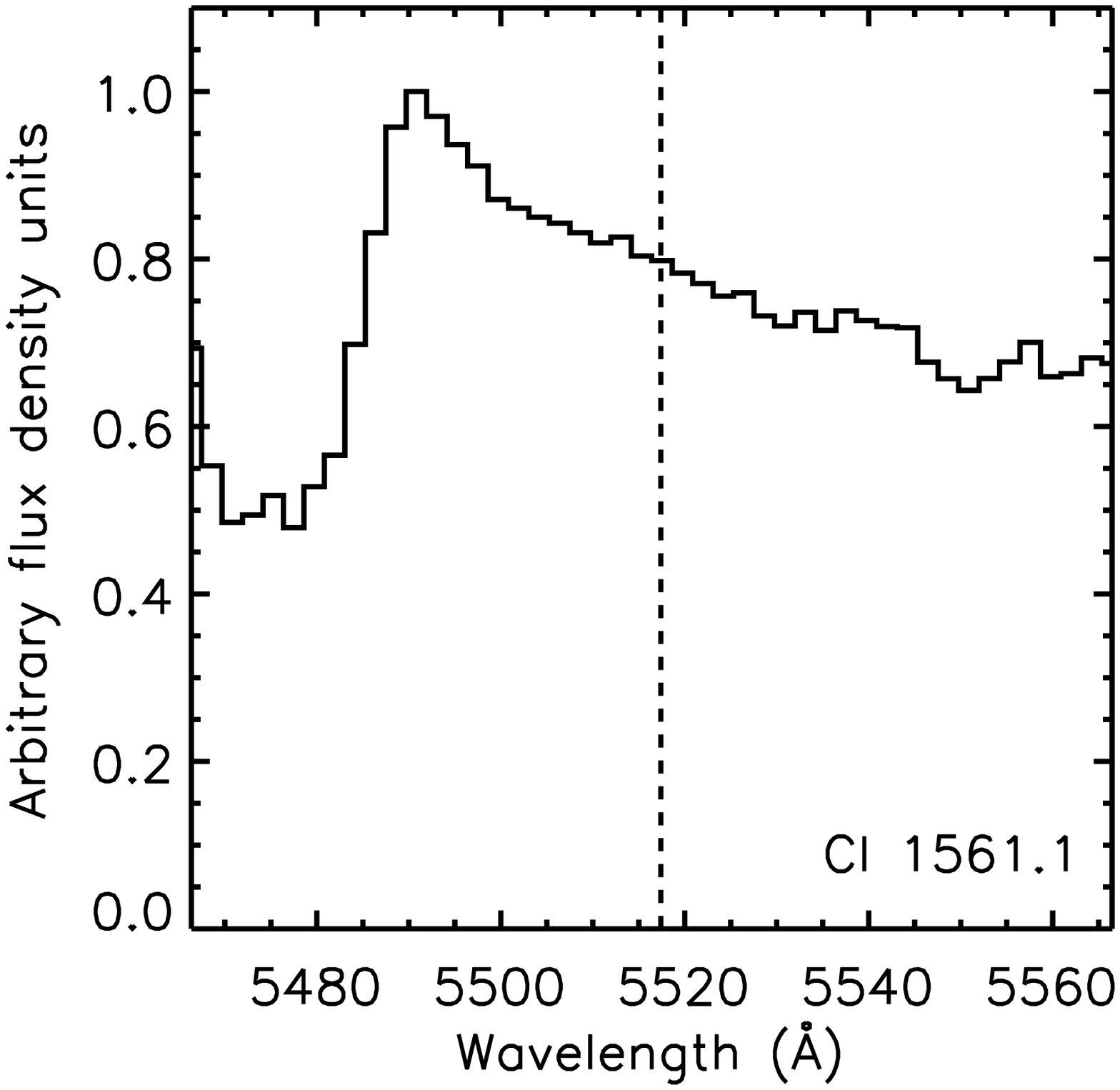}
\includegraphics{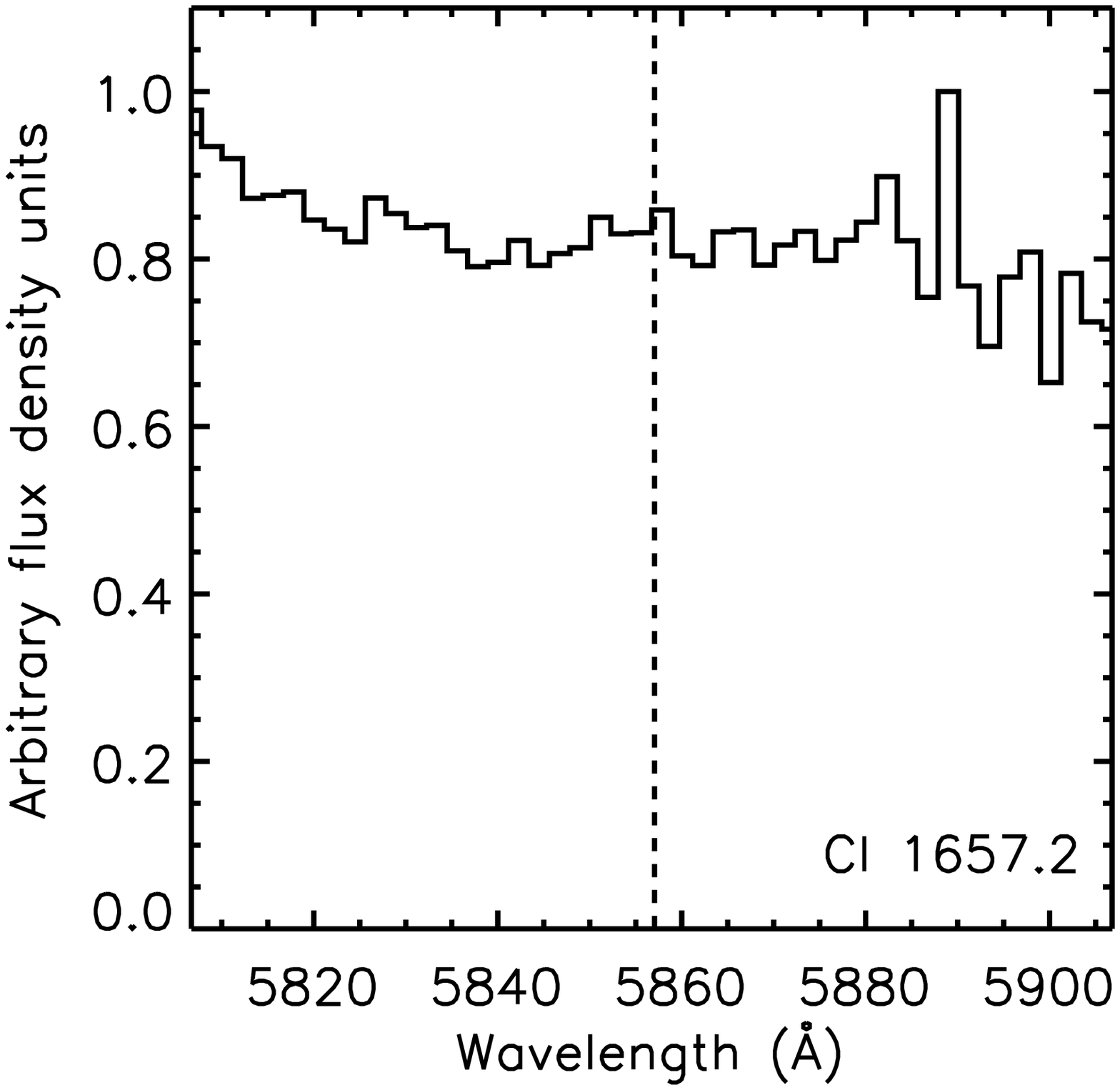}
\includegraphics{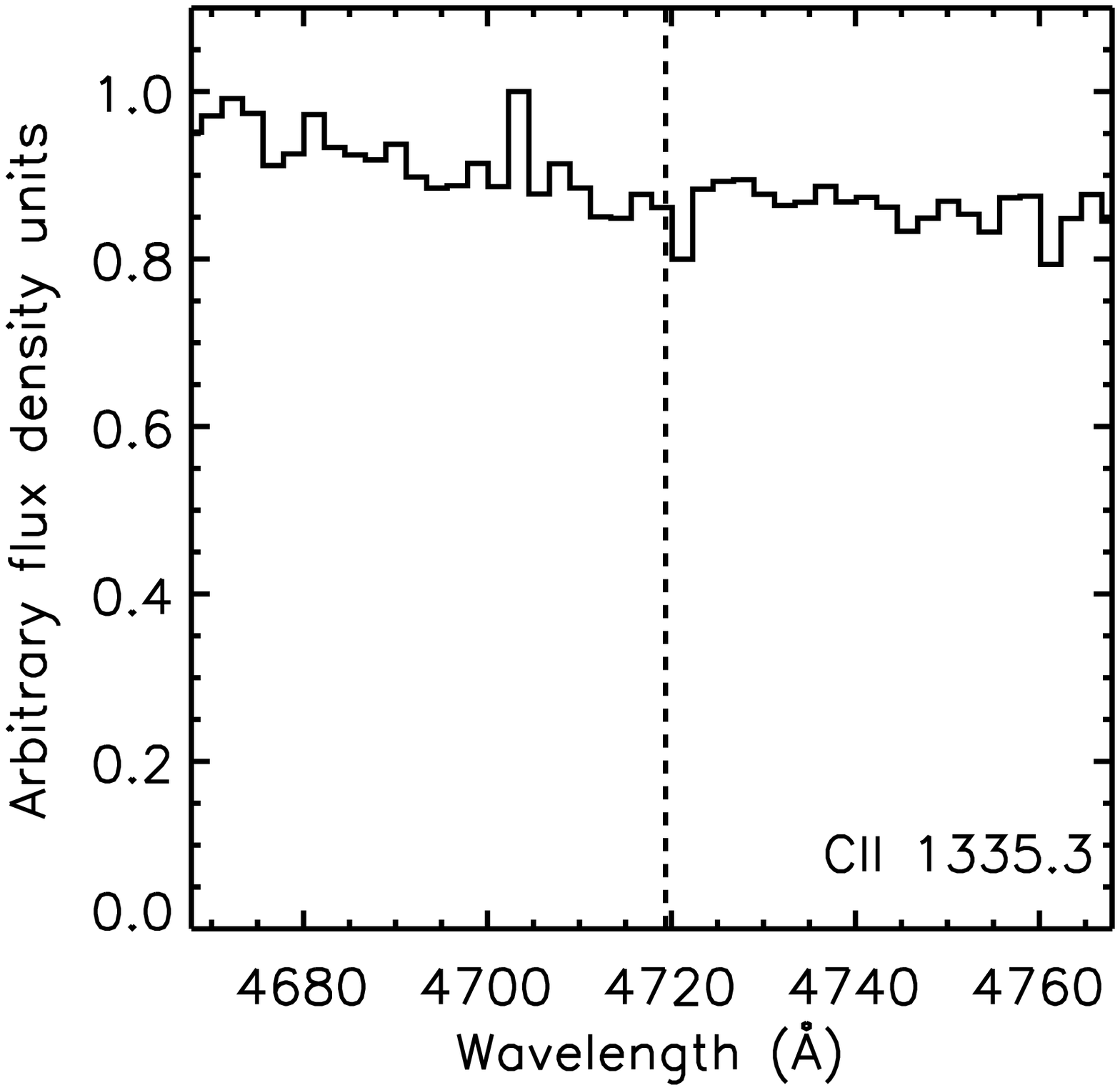}
\includegraphics{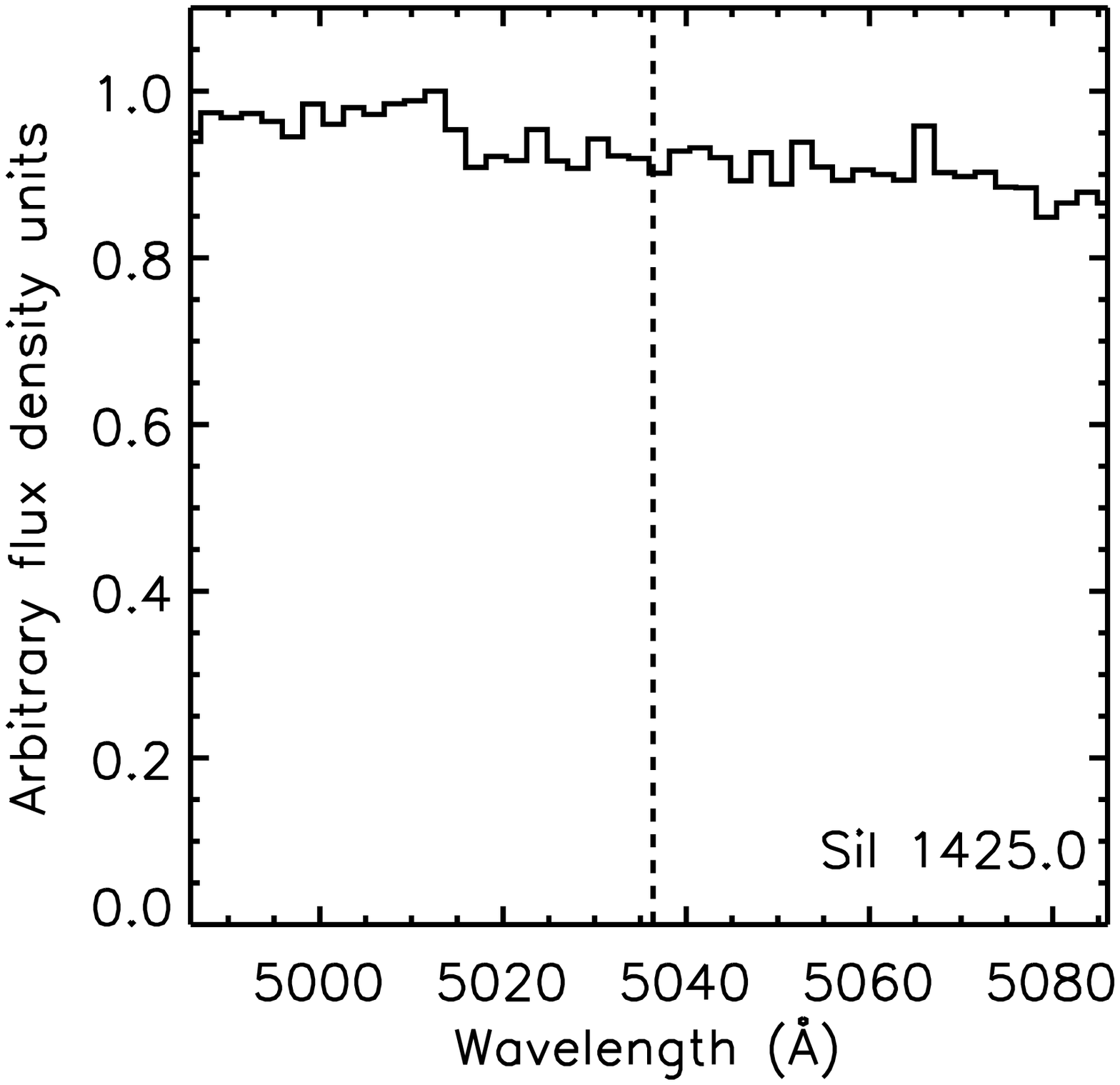}

\includegraphics{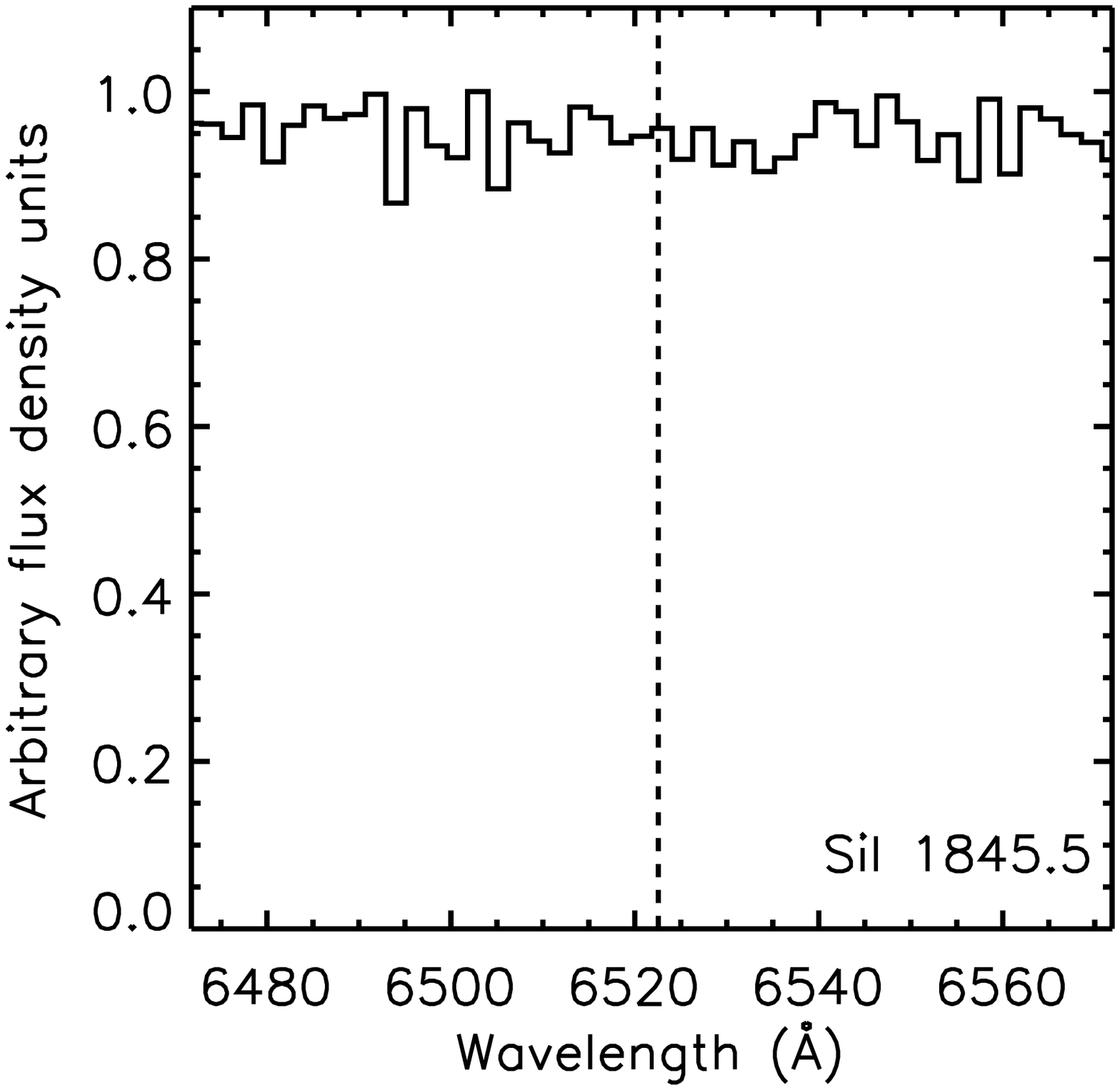}
\includegraphics{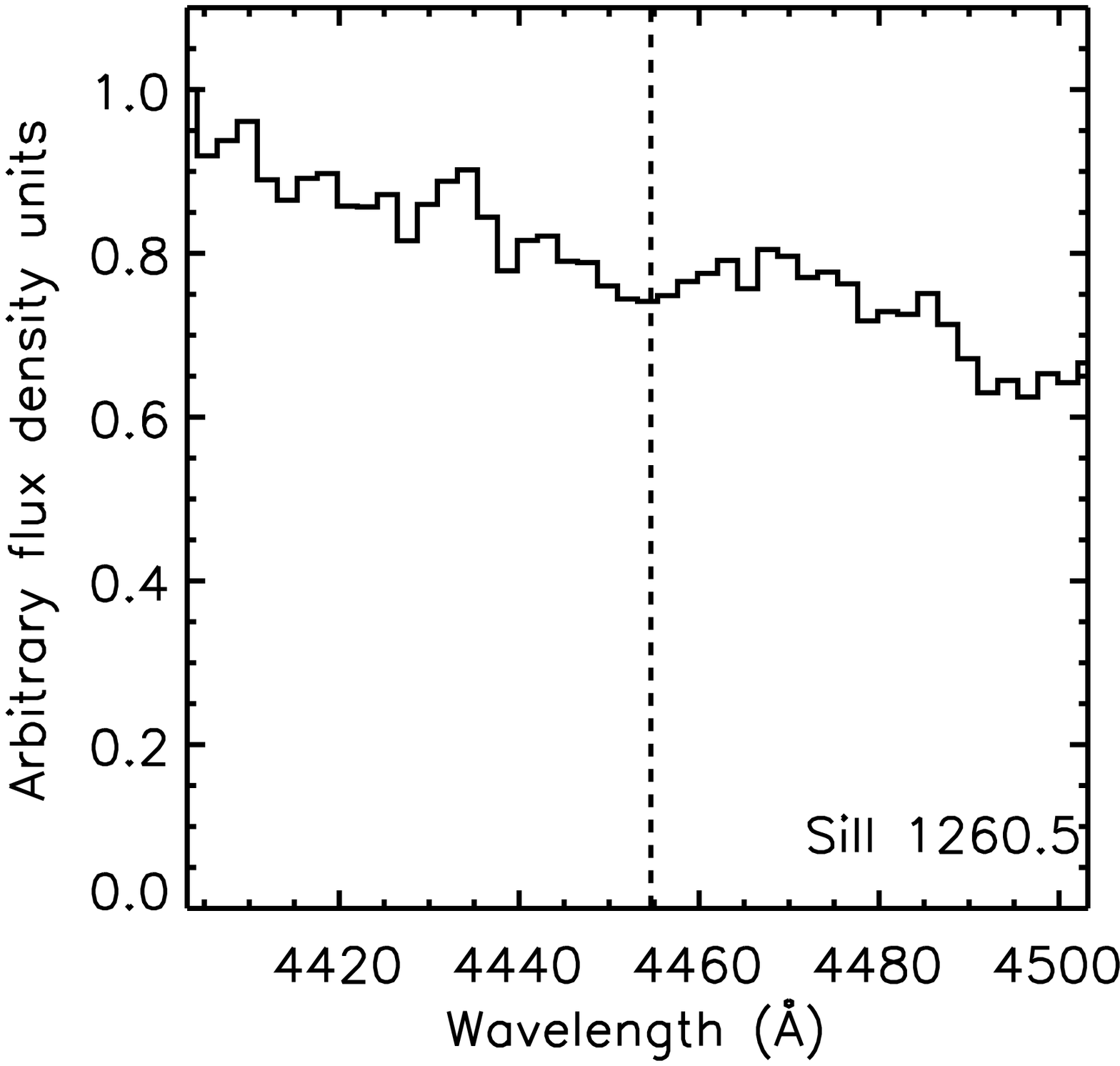}
\includegraphics{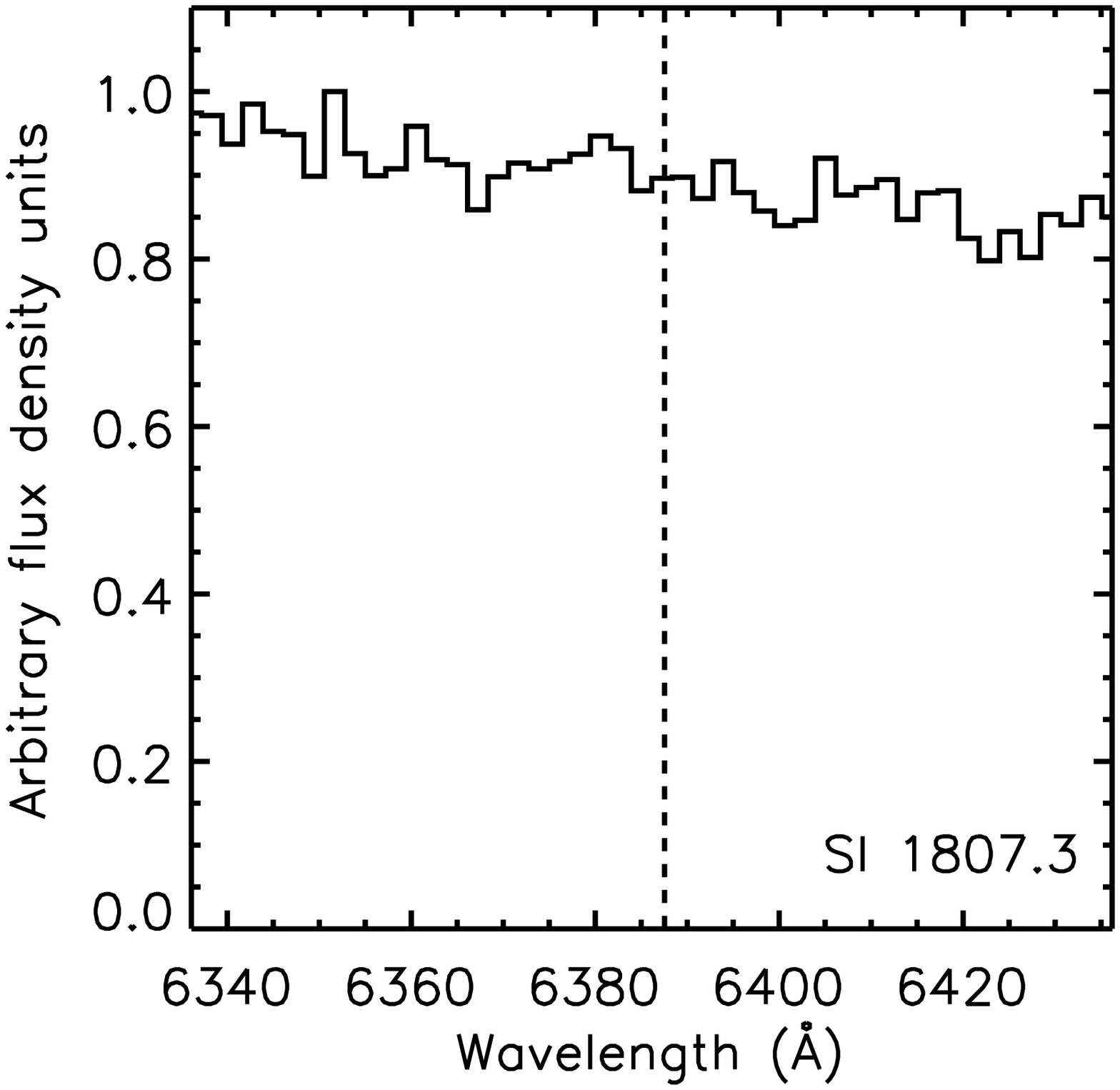}
\includegraphics{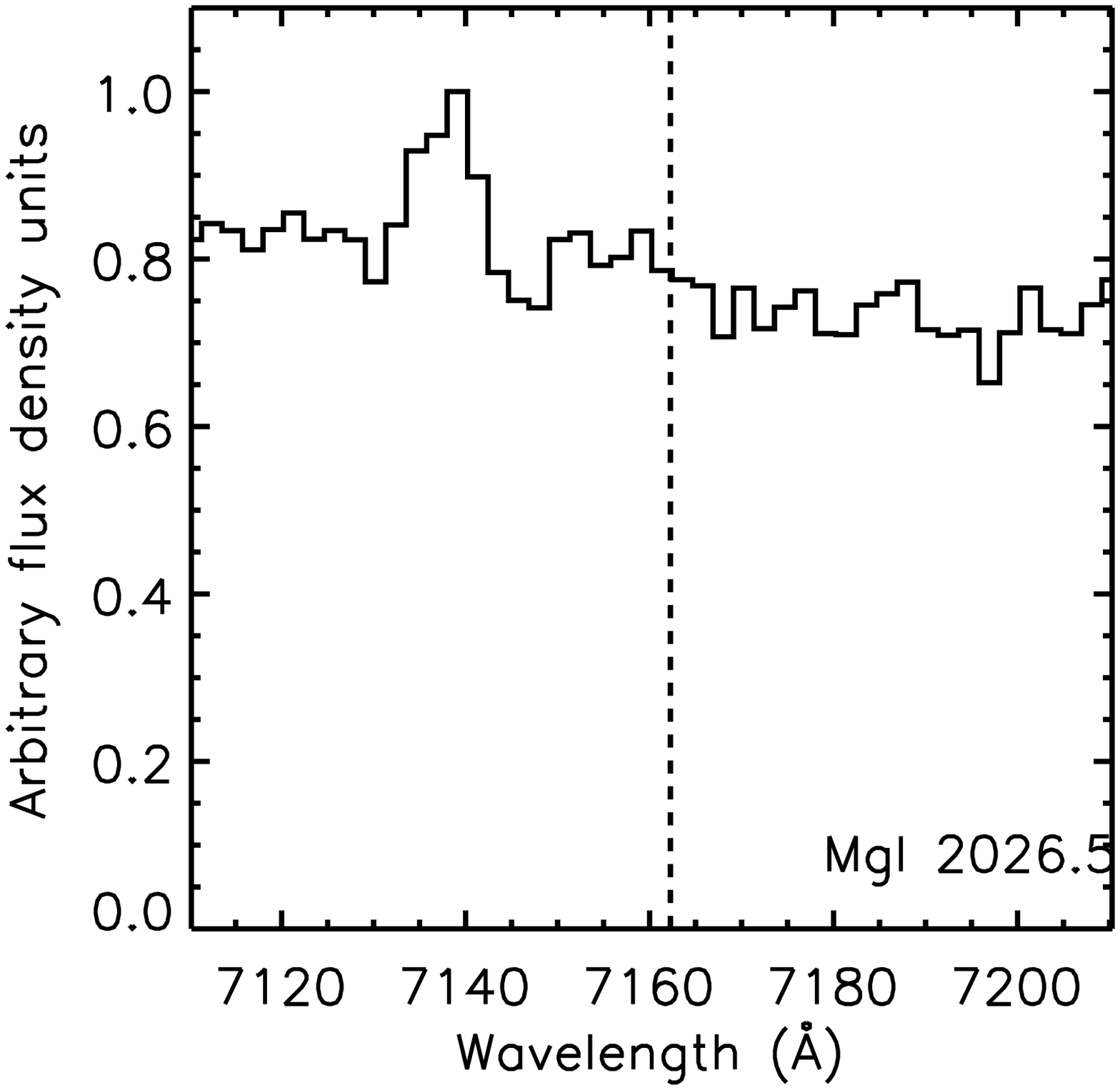}

\vspace{3.55in}
\caption{Sections of the one-dimensional spectrum of TXS 1436+157 showing the expected wavelengths of various undetected low-ionization lines.  The flux scale is arbitrary, and no smoothing has been performed.  Upper limits on $W_{\lambda}$ are given in Table \ref{columns}.}
\label{low-ionization}
\end{figure*}

Table \ref{columns} lists the detected absorption lines.  Velocities are given relative to the centroid of the narrow HeII $\lambda$1640 emission at the position of the quasar.  Also listed are upper limits for those absorption lines that are undetected in our data, but which are interesting in the context of the goals of our study.  Fig. \ref{spectra} shows regions of the extracted one-dimensional spectrum around the detected NV, SiIV and CIV absorption lines.  Table \ref{columns} also includes upper limits for several potentially useful absorption lines.  These limits were determined by calculating the equivalent width that would correspond to a gaussian absorption feature that has an FWHM equal to that of the appropriate instrumental profile (8 \AA), and a minimum that is below the continuum level by at least twice the local 1$\sigma$ pixel to pixel uncertainty in flux density.  The approximate expected wavelength of the OI $\lambda$1302.2 line coincides with a detected absorption line at $\sim$4606 \AA, but the presence of an absorption feature at this wavelength in the spectra of the two intermediate redshift star forming galaxies (Fig. \ref{OI_spectra}) suggests it may instead be due to an absorption system in front of all three galaxies (i.e., at z$\la$0.16). Therefore, we adopt the equivalent width of this aborption line as an upper limit to that of OI $\lambda$1302.2.  In Fig \ref{low-ionization} we illustrate the conspicuous absence of various other low-ionization absorption lines.

\subsection{Column Densities using the CoG}

We use the curve of growth (CoG) method to derive column densities from the observed values of $W_{\lambda}$ (Spitzer 1978).  First, we must ascertain on which part of the CoG are the absorption lines.  For the two clearly resolved doublets NV $\lambda\lambda$1239,1241 and SiIV $\lambda\lambda$1394,1403 the ratio between the equivalent width of the blue and red line, $W_{blue}$/$W_{red}$ can be used.  $W_{blue}$/$W_{red}$=2.0, 1.1, or $\sqrt{2}$, for the linear, flat, or damping parts of the CoG, respectively.  For the NV doublet, the $W_{blue}$/$W_{red}$=2.1$\pm$0.5 ratio we measure places it on the linear part of the CoG.  In the case of the SiIV doublet the measured $W_{blue}$/$W_{red}$=1.6$\pm$0.6 is ambiguous.  However, if we assume that the SiIV lines have a similar intrinsic width to the NV lines, then the fact that $W_{SiIV} \sim W_{NV}$ would place the SiIV lines on the linear part of the CoG also.  

For an absorption line on the linear part of the CoG, the column density N can be calculated using

\begin{equation}
N = \frac{1.13\times10^{20} W_{\lambda,0}}{f \lambda{_0}{^2}}
\end{equation}

\noindent where f is the oscillator strength, ${W_{\lambda,0}}$ is the rest-frame equivalent width (in \AA) and ${\lambda_0}$ is the rest-frame wavelength of the line (in \AA), and $N$ is in units of cm$^{-2}$.  The resulting $\sigma^{-1}$-weighted column densities are $N_{NV}$=2.45$\pm$0.23 $\times 10^{14}$ cm$^{-2}$ and $N_{SiIV}$=0.99$\pm$0.09 $\times 10^{14}$ cm$^{-2}$.  For CIV we apply this relation to our lower limit to $W_{CIV}$ to obtain a lower limit $N_{CIV}\ge$3.8 $\times 10^{14}$ cm$^{-2}$.

\begin{figure*}
\includegraphics{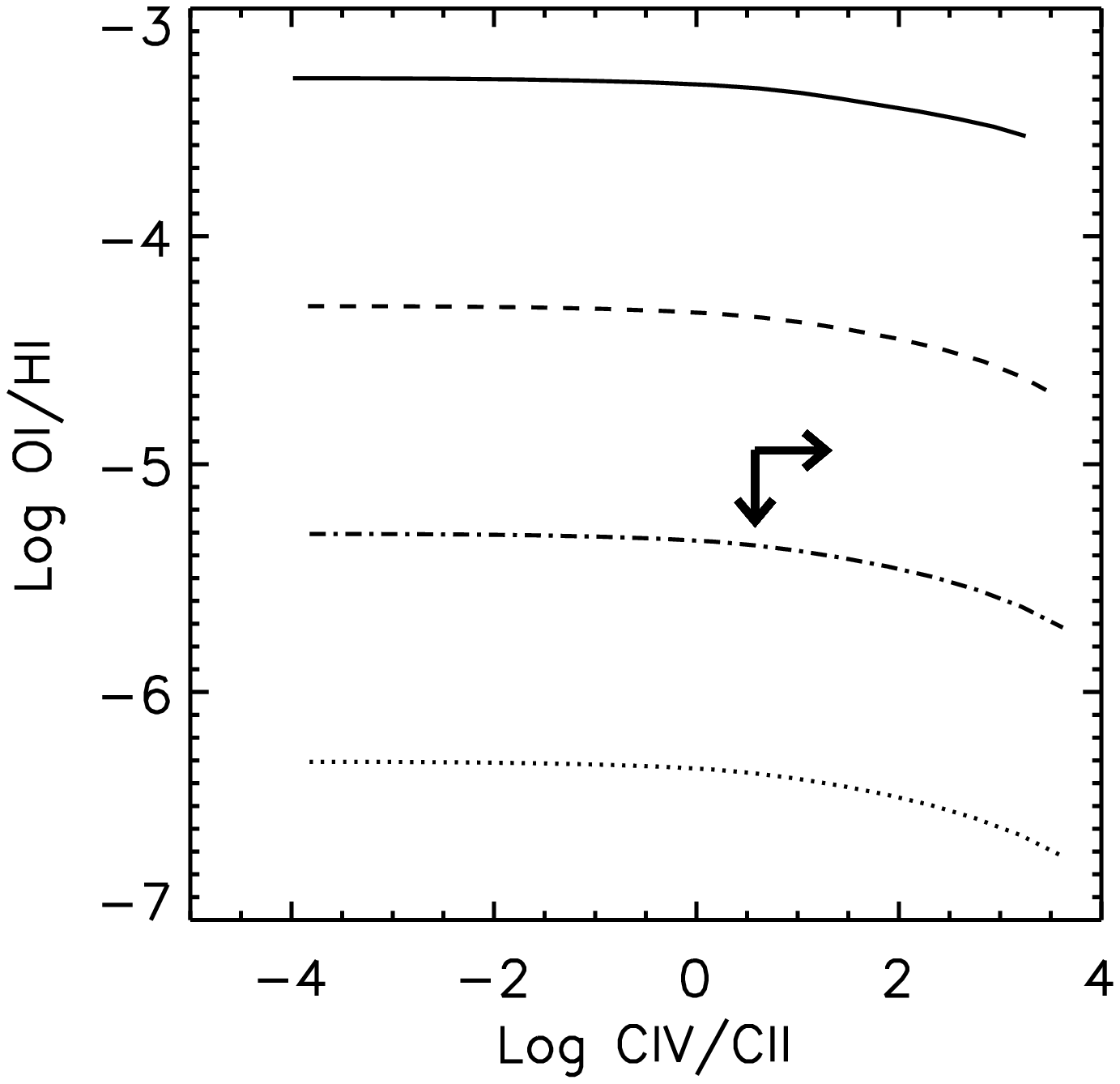}
\includegraphics{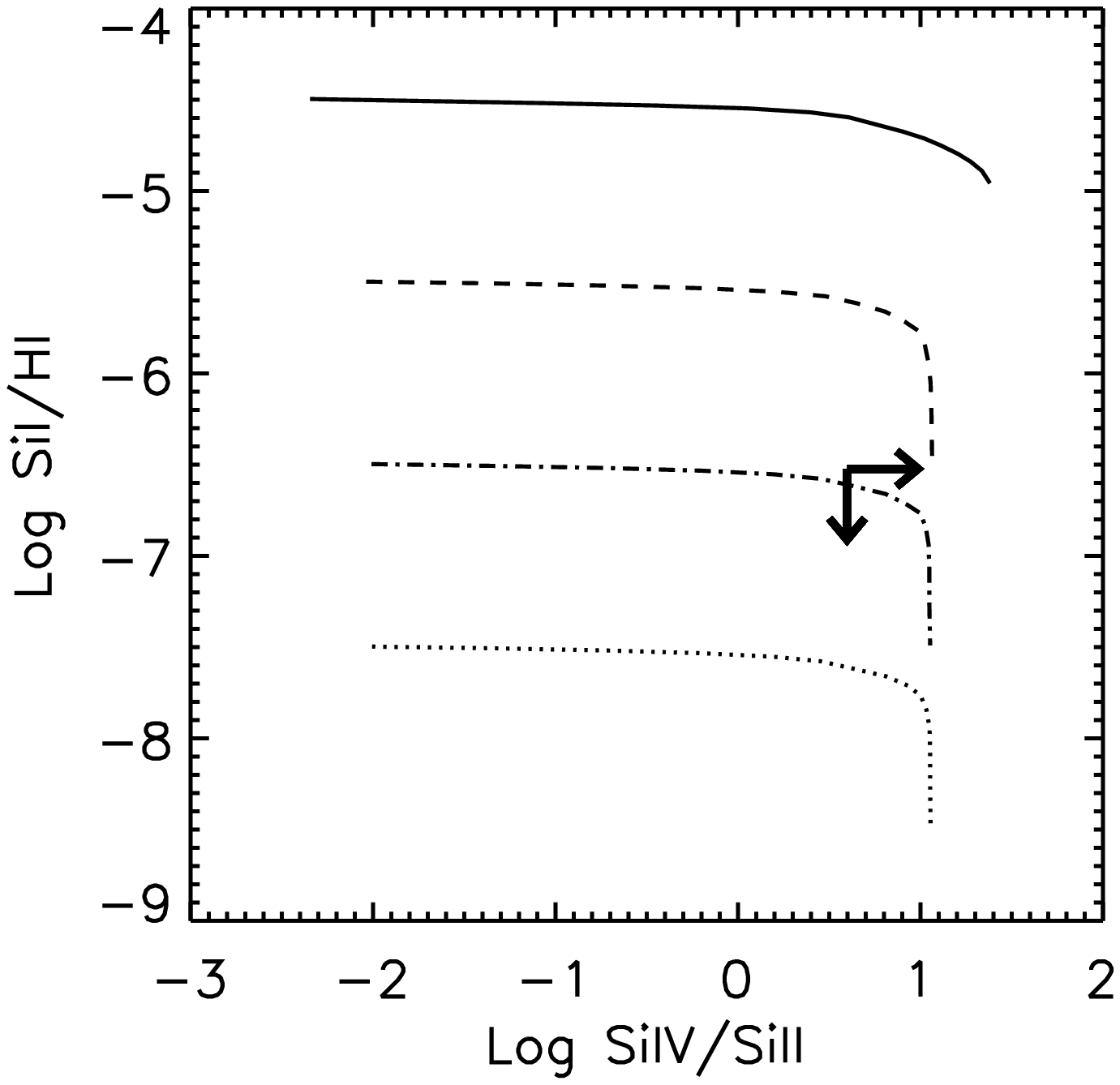}
\includegraphics{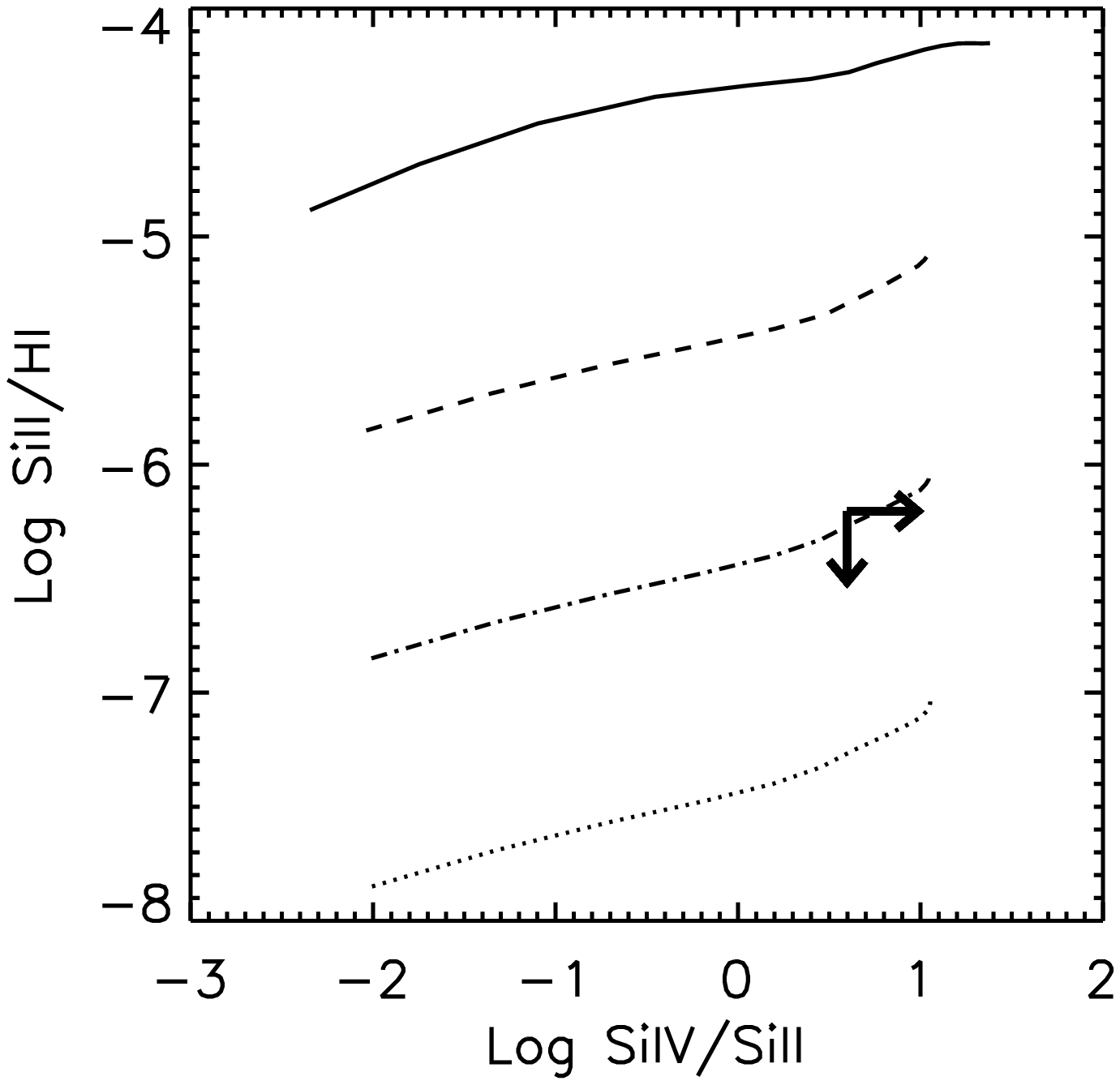}

\includegraphics{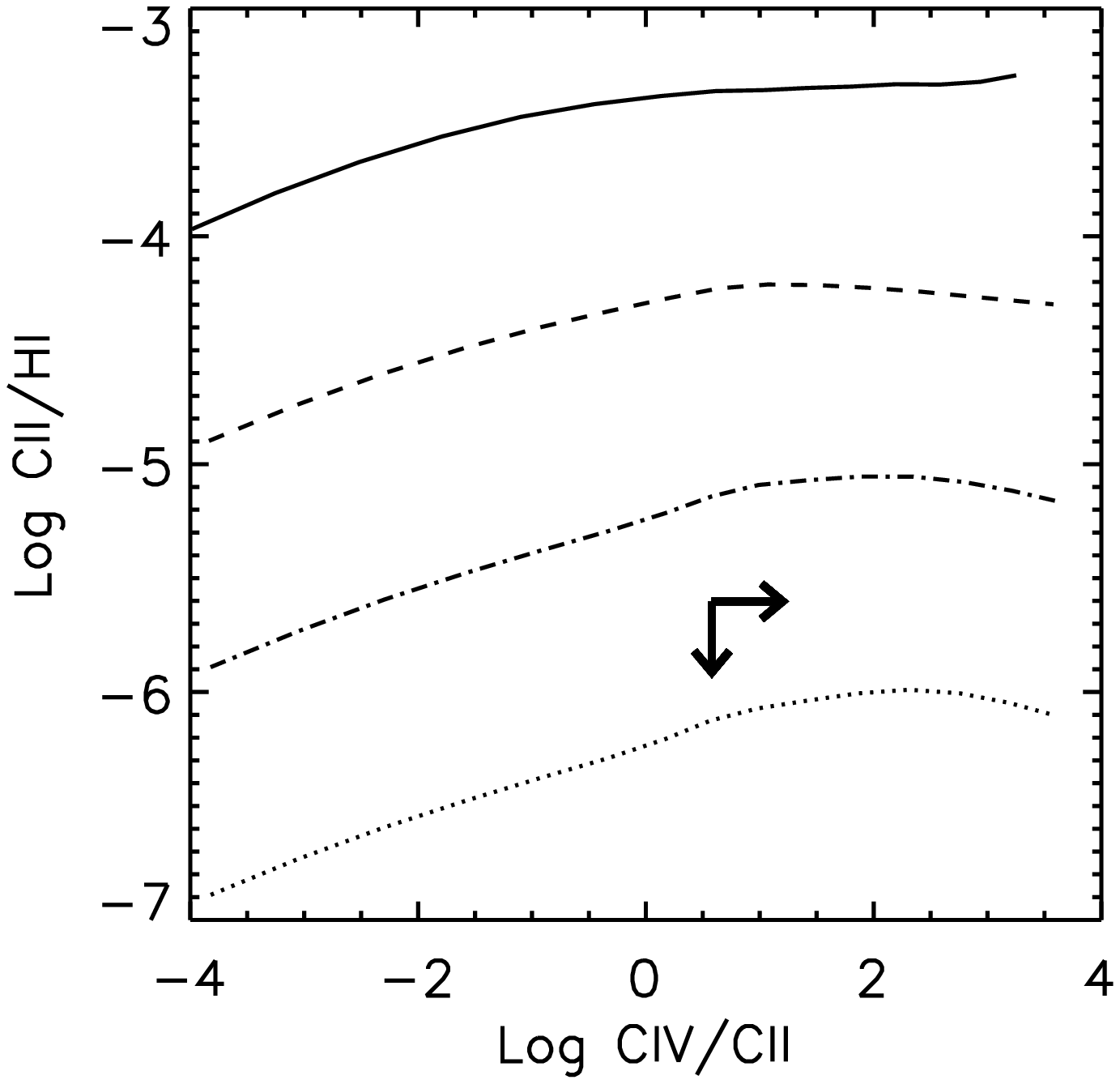}
\includegraphics{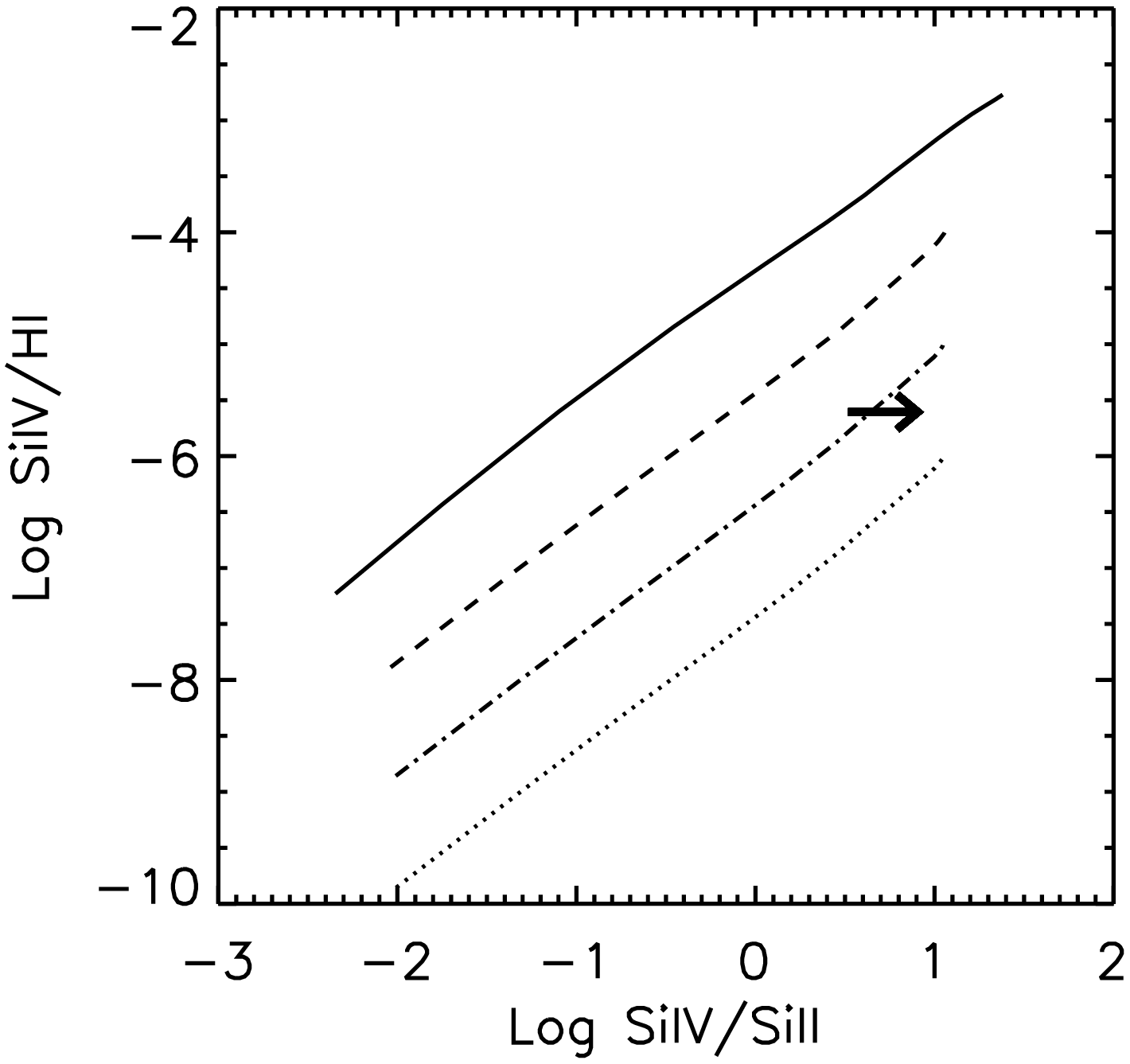}
\includegraphics{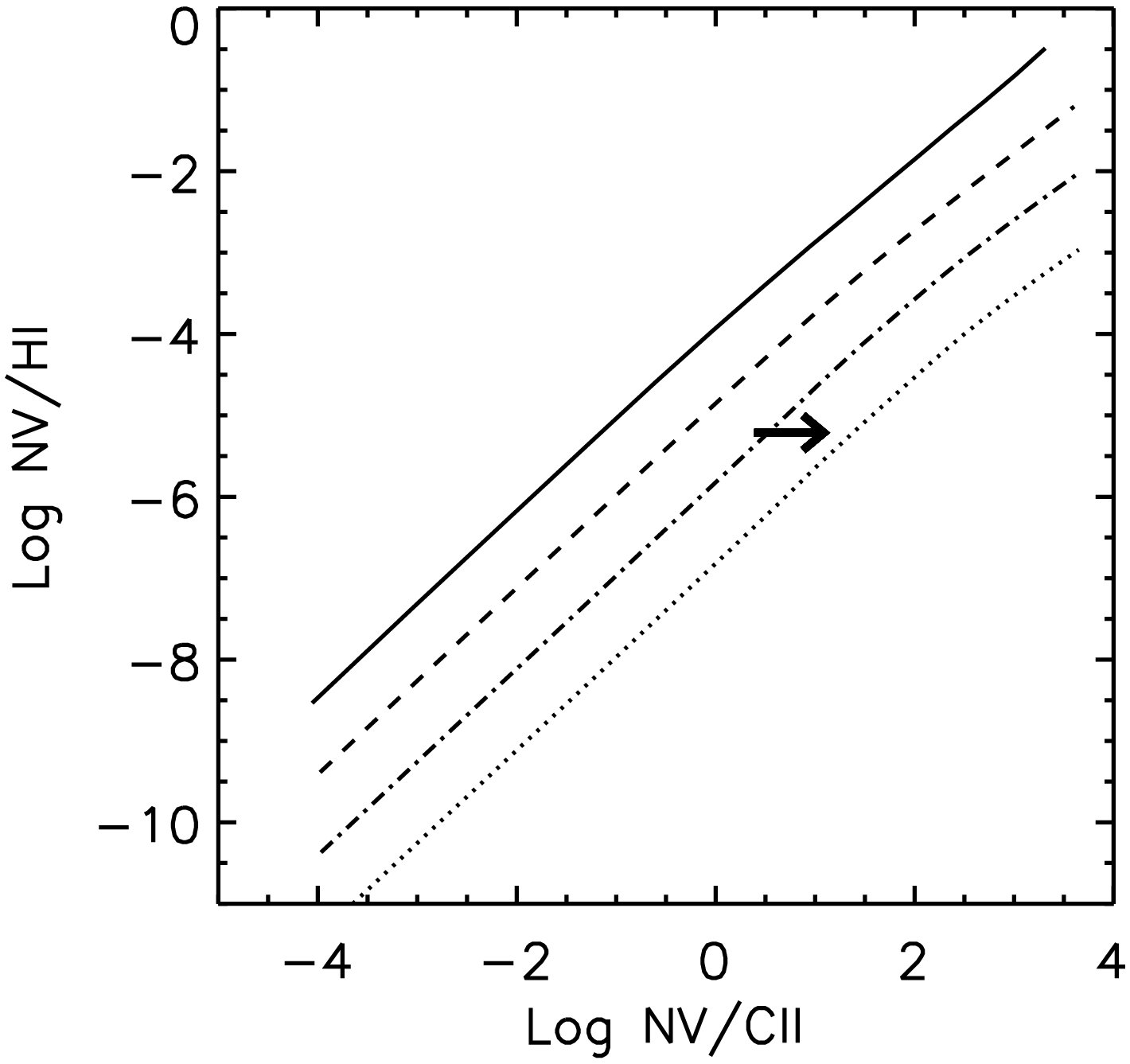}

\vspace{4.2in}
\caption{Photoionization model predictions for gas photoionized by an active galactic nucleus.  Each locus represents a sequence in U at four different metallicities (in the 12+log(O/H) system): 8.69 (solid line: solar value), 7.69 (dashed line), 6.69 (dash-dot line) and 5.59 (dotted line).  U increases from the left.  Arrows show the measurement or limits on the relevant column density ratios.}
\label{models1}
\end{figure*}

\subsection{Properties of the Absorber}
\subsubsection{Metallicity and ionization}
\label{abs_met_ioniz}
What can be said about the absorber independently of photoioniation models?  The detection of metal absorption lines from ionized species clearly demonstrates that the absorber is not composed of pristine material.  Moveover, it must be at least partially ionized, with the detection in absorption of the NV doublet signalling photoionization by a hard radiation field with a significant photon luminosity at $hv \ge 77.5$ eV ($\lambda \le$ 160 \AA).  Given that $N_{CIV}$ / $N_{CI}$ + $N_{CII} \ge$3.0 and $N_{SiIV}$ / $N_{SiI}$ + $N_{SiII} \ge$2.7, we can say that the absorber must be predominantly ionized.

In order to further investigate the ionization and chemical composition of the absorber, we have computed photoionization models using the code MAPPINGS Ic (Binette et al. 1985; Ferruit et al. 1997).  We adopt an ionizing spectral energy distribution appropriate for a powerful quasar, characterised by a power-law with spectral index $\alpha=-1.5$ (e.g. Robinson et al. 1987; this is justified by the detection of NV absorption lines).  The hydrogen density is set to $n_H=10$ cm$^{-3}$, but in the low-density regime, using a different value does not significantly affect the model results.  We adopt an isochoric behaviour for the gas density.  We take the ionization parameter\footnote{We define ionization parameter as $U=\frac{Q}{4 \pi r^2 c n_H}$, where $Q$ is the luminosity of ionizing photons of the ionizing source, r is the distance of the absorber from the ionizing source, and $N_H$ is the hydrogen density at the front face of the gas cloud.} U to vary from 0.0001 to 1.6 in steps of a factor of 2.  We adopt the solar abundances of Anders \& Grevesse (1989), except in the case of oxygen for which we adopt the more recent determination of O/H=4.9$\times 10^{-4}$ made by Allende Prieto, Lambert \& Asplund (2001).   We compute models for four different gas metallicities\footnote{We define metallicity as $Z$=12+log[O/H].}: $Z$=8.69 (solar), $Z$=7.69 (1/10 solar), $Z$=6.69 (1/100 solar) and $Z$=5.69 (1/1000 solar).  In the interest of simplicity, we scale all metals equally with oxygen.  Each model calculation is ended once the HI column density has reached 4$\times$10$^{19}$ cm$^{-2}$, to be consistent with the observations.  This means at low-U the models have a large neutral zone, while at the very high-U end the models become matter bounded.  

van Ojik et al. (1997) found that the absorbing gas is kinematically very quiescent, with a Doppler parameter of 37$\pm$20 km s$^{-1}$ measured from the Ly$\alpha$ absorption feature.  We have not considered ionization by shocks on account of this kinematic quiescence.  

Figure \ref{models1} shows various column density ratios resulting from our photoionization model calculations.  We have plotted various column density ratios involving high-ionization, low-ionization and neutral species in order to break the degeneracy between metallicity and ionization state.  Our lower limits on CIV/CII and SiIV/SiII require U$\ga$0.01 and $N_{H}\ga$10$^{21}$ cm$^{-2}$.   In addition, the position of the absorber relative to the model loci requires extremely low abundances of the relevant metals.  More specifically, $12+log[Si/H]$ and $12+log[N/H]$ are $\la$6.7 ($\la$1 per cent of solar), $12+log[C/H] \la$6.2 ($\la$ 0.3 per cent of solar).  

The OI/HI ratio gives us the slightly less sensitive limit $12+log[O/H] \la$7.3 ($\la$4 per cent solar).  It is interesting to note this is comparable to (or below) the oxygen abundance in the neutral phase in the halos of local star forming dwarf galaxies (e.g. Thuan et al. 2005), and also in the warm ionized phase of extremely metal-poor dwarf galaxies in the nearby Universe (e.g. Izotov et al. 1999; Papaderos et al. 2006; Izotov et al. 2006; Izotov et al. 2009).

\subsubsection{Geometry and Mass}
\label{mass}
In their moderate-resolution long-slit spectrum of TXS 1436+157, van Ojik et al. (1997) detected the HI Ly$\alpha$ absorption feature out to a projected radius of $\sim$40 kpc from the quasar.  This provides us with a rough lower limit of $\ga$40 kpc for the distance of the absorbing structure from the quasar.  If the absorbing gas is arranged in a spherical shell with radius r$\ge$40 kpc encircling TXS 1436+157, then its hydrogen mass can be estimated using

\begin{equation}
M_H = 4 \pi r^2 m_H N_H
\label{mass_eq}
\end{equation}

In this case the total mass of the absorber would be $\ge$1.6$\times 10^{11} M_{\odot}$.  If we assume instead that the absorber is a uniform sheet of gas then its mass would be

\begin{equation}
M_H = A m_H N_H
\end{equation}

\noindent where A is the total area of the absorber.  Assuming the absorber is as extended perpendicular to the radio axis as it is observed to be along the radio axis ($\sim$80 kpc: van Ojik et al. 1997), then its total hydrogen mass would be $\ge$5.1$\times 10^{10} M_{\odot}$.  

Since we have a lower limit on U ($\ga$0.01) and on the distance r between the quasar and the absorber ($\ge$40 kpc), and assuming the ionizing photon luminosity of the central AGN is $\le 10^{57}$ s$^{-1}$ (e.g. Villar-Mart\'{i}n et al. 2002), the density of the absorbing gas can then be constrained using

\begin{equation}
n_H=\frac{Q}{4 \pi r^2 c U}
\end{equation}

Under these assumptions, we obtain an upper limit of $n_H \le$ 18 cm$^{-3}$.  The physical thickness of the absorber along our line of sight toward the quasar is given by 

\begin{equation}
l=\frac{N_H}{n_H}
\end{equation}

\noindent which gives a lower limit of $l\ge$5.6$\times10^{19}$ cm, or $\ge$18 pc, assuming a volume filling factor of 1.  Significantly lower filling factors (originally suggested by van Ojik et al. 1997) do not seem plausible when one considers that the absorber has a covering factor close to unity over several tens of square kiloparsecs.  

It is also interesting to estimate the mass of metals contained within the absorber.  In our U$\ga$0.1 models CIV, NV and SiIV all have an ionization fraction on the order of $\sim$0.1, meaning that the total columns of C, N and Si are probably $\ga$3.8, $\sim$2.5 and $\sim$1.0 $\times$ 10$^{14}$ cm$^{-2}$, respectively.  Using equation \ref{mass_eq}, we then obtain C, N and Si masses of $\ga$6, $\sim$4 and $\sim$2 $\times$10$^{5}$ $M_{\odot}$, respectively.  If we adopt the solar Si/O abundance ratio, the mass of oxygen would then be $\sim$2 $\times$10$^{6}$ $M_{\odot}$.  According to the calculations of Moll\'a and Terlevich (2012), the production and injection of this mass of oxygen would require the formation of $\sim$10$^9$ $M_{\odot}$ of stars.  

\begin{figure*}
\includegraphics{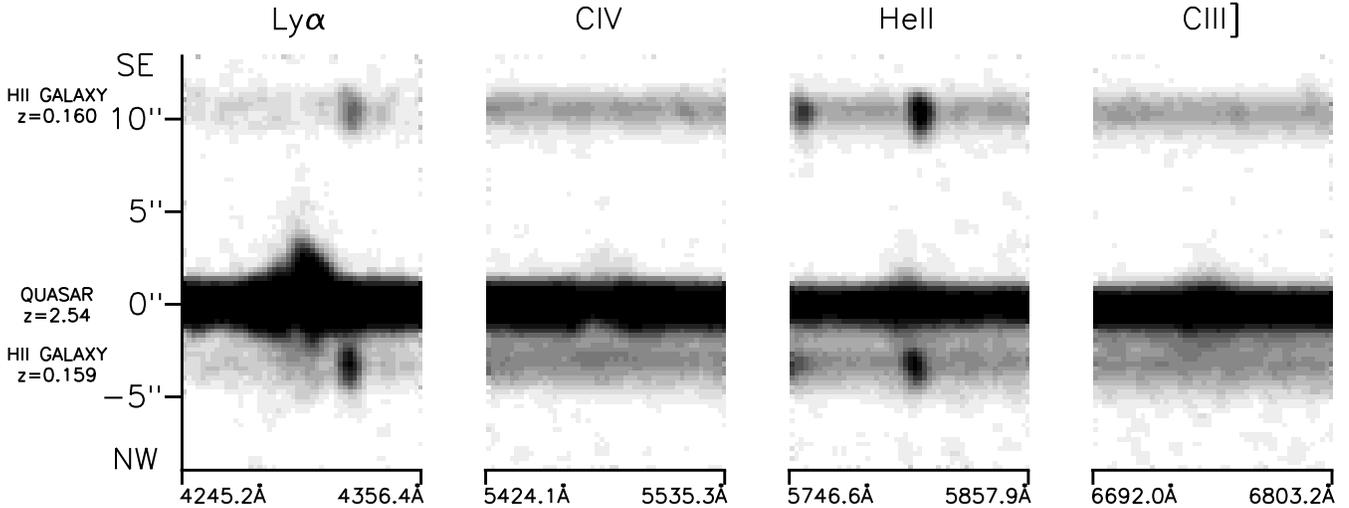}
\vspace{2.85in}
\caption{Sections of the two-dimensional spectrum of TXS 1436+157, centred on the brightest narrow emission lines: Ly$\alpha$, CIV, HeII and CIII].  The vertical scale is in arc seconds, while the horizontal scale gives observed-frame wavelength, in Angstroms.  Cut levels are chosen to show the spatially extended, narrow emission lines, and differ between panels.  The data have been smoothed by a 3 pixel box-car average to emphasize the extended line emission.}
\label{2d}
\end{figure*}

\begin{table}
\centering
\caption{Emission line ratios measured in the extended line emitting nebula 2.0\arcsec - 3.8\arcsec East of the quasar.  For comparison, we also show the values of the ratios from the z$>$2 radio galaxy (HzRG) average emission line spectrum of Humphrey et al. (2008b).} 
\begin{tabular}{lll}
\hline
Ratio & Value & HzRG average \\
\hline
Ly$\alpha$ / HeII $\lambda$1640 & 22$\pm$4 & 11.7 \\
NV $\lambda$1241 / HeII $\lambda$1640 & $\le$0.3 & 0.3\\
NV $\lambda$1241 / CIV $\lambda$1549 & $\le$0.2 & 0.5 \\
NV $\lambda$1241 / Ly$\alpha$ & $\le$0.014 & 0.04 \\
CIV $\lambda$1549 / HeII $\lambda$1640 & 1.3$\pm$0.2 & 1.8 \\
CIV $\lambda$1549 / CIII] $\lambda$1908 & 4.2$\pm$1.5 & 2.4 \\
CIII] $\lambda$1908 / HeII $\lambda$1640 & 0.3$\pm$0.1 & 0.73 \\
\hline
\end{tabular}
\label{ratios}
\end{table}

\subsection{Detectability of absorbing shells in Ly$\alpha$ emission}
Would an absorber of the kind associated with TXS 1436+157 be detectable in emission?  For a quasar ionizing luminosity Q$\sim 10^{57}$ s$^{-1}$, a torus opening angle of $\sim$ 90$^{\circ}$, shell radius $\sim$40 kpc and covering factor $\sim$1, then the Ly$\alpha$ luminosity of the absorbing shell would be $\sim$3$\times$10$^{45}$ erg s$^{-1}$, comparable to the Ly$\alpha$ luminosity of the extended nebulae photoionized by quasars and radio galaxies.  This assumes intervening gas in the host galaxy has a low covering factor and that the shell is thick enough to absorb all incident ionizing photons.  In narrow band imaging, such a shell would appear as a pair of diametrically opposed arcs centred on the quasar, each with a Ly$\alpha$ surface brightness on the order of $\sim$10$^{-15}$ erg s$^{-1}$ cm$^{-2}$ arcsec$^{-2}$.  Thus, under the right conditions -- a luminous enough quasar surrounded by a shell of sufficient thickness and covering factor -- a gaseous structure of this kind should indeed be detectable in Ly$\alpha$ emission.  

Interestingly, van Ojik et al. (1997) found that the Ly$\alpha$ absorption feature of TXS 1436+157 is not black at its trough, with a residual flux of $\sim$10$^{-16}$ erg s$^{-1}$ in that spectral region.  This result is qualitatively consistent with the idea of some {\it in situ} production of Ly$\alpha$ photons in the absorber, due to its photoionization by the active nucleus, although other possible explanations exist, such as a covering factor of $<$1.

\subsection{Comparison against other extended metal line absorbers}
In this section we compare and contrast the properties of the absorber associated with TXS 1436+157 the small number of other spatially extended metal line absorbers known to be associated with high-z active galaxies.  

A similar analysis to that we have performed has been done for another radio loud quasar, MRC 2025-218 at z=2.63 which, like TXS 1436+157, shows a spatially extended Ly$\alpha$ absorber seen in front of the galaxy's quasar-photoionized LAN (Humphrey et al. 2008a).  In Table \ref{mrc2025} we reproduce from Humphrey et al. (2008a) the properties of that absorber.  The absorbers of MRC 2025-218 and TXS 1436+157 show remarkably similar NV column densities, which are 3.6$\pm$1.4$\times 10^{14}$ and 2.5$\pm$0.2$\times 10^{14}$ cm$^{-2}$, respectively, which itself suggests the two absorber systems have surprisingly similar (total) nitrogen column densities, and a fairly similar $U=\frac{Q}{4\pi r^2 c n_H}$.  Moreover, both absorbers are predominantly ionized as shown by their large N$_{SiIV}$/N$_{SiI}$ and N$_{CIV}$/N$_{CI}$ ratios.  The limits on their gas density, although pulling in opposite directions, nonetheless permit a similar density of $\sim$10 cm$^{-3}$.  

A comparison with the spatially extended metal line absorbers associated with high-z radio galaxies (z$\ga$2: HzRGs) is also interesting.  While many HzRGs show spatially extended HI absorption in front of their extended LAN (e.g. van Ojik et al. 1997), only two have also been unambiguously detected in metal lines.  In the case of the z=2.92 HzRG MRC 0943-242, Binette et al. (2000) have detected CIV absorption at the redshift of the extended HI absorber, while Jarvis et al. (2003) detected in CIV absorption the extended HI absorber of the z=2.23 HzRG MRC 0200+015.  Interestingly, the two HzRGs have almost identical values of N$_{CIV}$: 3.8$\pm$0.4$\times 10^{14}$ and 4.9$\pm$0.2$\times 10^{14}$ cm$^{-2}$ for MRC 0943-242 and MRC 0200+105, respectively, which suggests their absorbers have similar (total) carbon column densities and similar $U=\frac{Q}{4\pi r^2 c n_H}$.  Compared to these values, the extended absorber in front of MRC 2025-218 has a substantially higher CIV column density of N$_{CIV}$=20$^{+20}_{-10} \times 10^{14}$ cm$^{-2}$, while the absorber in front of TXS 1436+157 has N$_{CIV} \ge$3.8$\times 10^{14}$ cm$^{-2}$ which is about as high as, or higher than the two HzRGs.  This relative enhancement of highly-ionized species in the extended absorbers associated with the quasars is consistent with the idea that those absorbers are photoionized by a more intense radiation field -- presumably that of the quasar.  

\begin{table}
\centering
\caption{Properties of the associated absorber in front of the z=2.63 radio loud quasar MRC 2025-218, reproduced from Humphrey et al. (2008a).  $\Delta v$ is the velocity shift of the absorber from the velocity centroid of the narrow HeII $\lambda$1640 emission line measured at the projected position of the quasar.} 
\begin{tabular}{ll}
\hline
Species   & $N$ (10$^{14}$ cm$^{-2}$)    \\  
\hline
HI         & $\ga$300    \\
CI         & 1.4$\pm$0.3 \\
CII        & $\ge$9.9  \\
CIV        & 20$^{+20}_{-10}$ \\  
NV         & 3.6$\pm$1.4 \\
OI         & $\ge$6.8    \\
SiI        & $\le$0.37   \\
SiII       & 13$\pm$11  \\
SiII*      & 1.7$\pm$0.7 \\
SiIV       & $\ge$2.6   \\
AlII       & $\ge$0.16  \\
AlIII      & 0.4$\pm$0.1  \\
\hline
Property  & Value \\
\hline
r          & $\ge$25 kpc \\
$M_H$        & $\ga 10^{8} M_{\odot}$ \\
$n_H$      & $\ge$10 cm$^{-3}$ \\
$\Delta v$ & -600 km s$^{-1}$\\
\hline
\end{tabular}
\label{mrc2025}
\end{table}

\subsection{The Ly$\alpha$  nebula}
The properties of the extended Ly$\alpha$ emitting nebula (LAN) are qualitatively consistent with the results presented by van Ojik et al. (1997).  The extended Ly$\alpha$ emission has FWHM=820$\pm$50 km s$^{-1}$, shows an excess of flux in its blue wing, and is red-shifted by 500 km s$^{-1}$ relative to the HeII emission at the position of the quasar: these properties are consistent with effects expected due to the strong, spatially extended Ly$\alpha$ absorption on the blue side of the line profile.   Our new results are as follows.  

\subsubsection{Kinematic properties: infall}
\label{infall_section}
Out of the UV emission lines detected in the spectrum of TXS 1436+157, the HeII $\lambda$1640 line is expected to give the most reliable indication of the kinematic properties of the gas in the LAN.  As a non-resonant recombination line from helium its emissivity should be the least sensitive overall to absorption, geometry, metallicity or ionization parameter.  In the Eastern region of the LAN the HeII line shows a significant red-shift of 110 km s$^{-1}$ from the position of the quasar to $\ga$1\arcsec to the East\footnote{On the West side of the quasar HeII emission was not detected, although this may simply be due to obscuration by the intervening galaxy at z=0.159.}, and has FWHM=460$\pm$40 km s$^{-1}$.  These properties are consistent with gravitational motion within a massive galaxy and its dark matter halo.  The quiescence of the kinematics are surprising given the small size of the radio source (45 kpc: Carilli et al. 1997), which would usually be associated with extreme gas kinematics (FWHM$\ga$1000 km s$^{-1}$: e.g., Best, R\"ottgering \& Longair 2000; Humphrey et al. 2006; Nesvadba et al. 2006).  In order to better understand the kinematic properties of the HeII line, we must take into consideration orientation effects.  

In the standard paradigm for powerful active galaxies, the ionizing radiation of the quasar is collimated into two diametrically opposed ’ionization cones’ by an optically thick toroidal structure around the nucleus.  In the case of TXS 1436+157, our relatively direct view of the quasar then shows we are looking through the base of one of the ionization cones.  Gas infalling into the host galaxy is caught in the ionizing radiation field of the quasar, and reprocesses ionizing photons into narrow emission lines.  Along the line of sight to the quasar, relatively bright line emission from ionized gas which is relatively closer to the quasar, and which is nearer to the systemic velocity of the galaxy, should dominate over low-surface brightness line emission from infalling gas. However, along lines of sight substantially offset from the position of the quasar, the line emission would be dominated by the infalling gas and will show a relative red-shift.  If we assume that the HeII line at the position of the quasar is emitted close to the systemic velocity, then the relative red-shift of the line in the Eastern LAN indicates that the nebula is in inflow/infall into the host galaxy.  This is consistent with previous studies which have concluded that the Ly$\alpha$ nebulae around high-z radio galaxies and quasars often have a significant infalling/inflowing component (Weidinger, M{\o}ller \& Fynbo 2004; Humphrey et al. 2007; Villar-Mart\'{i}n et al. 2007a).  
 
Only a very rough estimate of the gas infall rate can be obtained from the available information.  If we adopt a gas density of $\sim$50 cm$^{-2}$, a filling factor $\sim$10$^{-5}$ (e.g. McCarthy 1993; Villar-Mart\'{i}n et al. 2003), and an infall velocity of 110 km s$^{-1}$ as implied by the velocity red-shift of the HeII line from the position of the quasar to the East region of the nebula, then the infall rate would be $\sim$10-100 $M_{\odot}$ yr$^{-1}$.  This is comparable to the gas infall rates estimated by Humphrey et al. (2007) for the giant Ly$\alpha$ nebulae of a sample of radio galaxies at z$>$2, and is sufficient to power a quasar at the top of the luminosity function ($L_{bol}\sim$10$^{46}$ erg s$^{-1}$).

\subsubsection{Excitation: Ly$\alpha$ excess}
To measure representative line ratios for the LAN, from our coadded two dimentional spectrum (shown in Fig \ref{2d}) we have collapsed spatially a one-dimensional spectrum from regions at projected distances 2.0\arcsec - 3.8\arcsec East from the position of the quasar.  The measurements were done by integrating the total flux above the continuum level, with 1$\sigma$ uncertainties dominated by the uncertainty in the continuum level.  The line ratios are listed in Table \ref{ratios}.  The three line ratios formed from only CIV $\lambda$1549, HeII $\lambda$1640 and/or CIII] $\lambda$1908 are fairly consistent with the z$>$2 radio galaxy average (Humphrey et al. 2008b; see also McCarthy 1993).  The detection of the HeII line shows that the LAN is ionized at least in part by the hard radiation field of the quasar, while the detection of the two carbon lines indicate that the gas is enriched in metals.  As discussed by Humphrey et al. (2008b) and others, line ratios using only CIV $\lambda$1549, HeII $\lambda$1640 and CIII] $\lambda$1908 do not provide strong constraints on the metallicity of the emitting gas: metallicities as low as $\sim$5 per cent solar, or as high as a few times solar, are possible for the LAN.  However, the non-detection of narrow NV emission argues against very high metallicities (see e.g. Vernet et al. 2001; Humphrey et al. 2008b).  

The Ly$\alpha$/HeII $\lambda$1640 ratio, on the other hand, is a factor of $\sim$2 higher than the average value, and is significantly higher than the range of values $\sim$8-15 that `normal' AGN-photoionization models can produce (Humphrey et al. 2008b).  Judging from the fitting carried out by van Ojik et al. (1997) for the extended Ly$\alpha$ emission, the pre-absorbed Ly$\alpha$ flux ought to be a factor of about $\times$1.3 higher, which would raise the Ly$\alpha$/HeII ratio to $\sim$29.  Ly$\alpha$ excesses of this kind have previously been identified in the LANs of a number of z$>$2 radio galaxies (e.g. van Ojik et al. 1994; Villar-Mart\'{i}n et al. 2007b), with the previously proposed explanations being (a) cooling radiation from low-metallicity infalling gas; (b) higher than expected nebular temperatures, due to low gas metallicity in the photoionized nebulae; (c) resonance flourescence; or (d) photoionization by young stars (see e.g. Villar-Mart\'{i}n et al. 2007b).  In the case of the Ly$\alpha$ nebula around TXS 1436+157, however, we must discard options (a) and (b) due to the substantial level of chemical enrichment of the nebula, so that only (c) and (d) remain as plausible explanations.  Measuring the Ly$\alpha$/H$\alpha$ ratio (e.g. Hayashi et al. 2012), or polarimetry of the Ly$\alpha$ emission (e.g. Hayes, Scarlata \& Siana 2011), would be useful to distinguish between these possibilities.  

\subsection{The Nature of the Absorber and Ly$\alpha$ nebula}
\label{scenario}
Assuming that the narrow HeII emission at the position of the quasar is emitted close to the systemic velocity, then the relative blueshift of the absorber implies it is in outflow.   Various previous investigators have discussed the possible origins of similar blue-shifted, absorbing structures around high-z radio galaxies and quasars (e.g. van Ojik et al. 1997; Binette et al. 2000; Krause 2002; Jarvis et al. 2003; Wilman et al. 2004; Humphrey et al. 2008a).  In the case of TXS 1436+157, we favour a scenario adapted from the super-bubble scenario devised by Tenorio-Tagle et al. (1999).  During a powerful starburst, mechanical energy is deposited by supernovae explosions into the galaxian interstellar medium, producing in an expanding superbubble which decelerates as it sweeps up gas from the interstellar medium.  The shell becomes Rayleigh-Taylor unstable and fragments (blows-out), thereby allowing the thermalized supernovae ejecta to rapidly escape into the low density halo of the galaxy, which then forms a new expanding shell sweeping up low metallicity gas from the halo.  Left behind from the fragmentation of the initial shell are Rayleigh-Taylor 'fingers' of cooler, denser, metal-enriched gas.  These take $\sim$100 Myr to fall from a radius of the order of $\sim$10 kpc into the central regions of the galaxy, at a velocity of $\sim$100 km s$^{-1}$ (see $\S$\ref{infall_section}), to provide fuel to trigger the quasar and radio jets.  The radiation field produced by the quasar then photoionizes the infalling material, thereby allowing its detection in Ly$\alpha$ and other lines, and also provides a background source for detection of the new shell, which has now expanded to a radius of several tens of kpc during the delay between the blow-out event and the triggering of the quasar.  

Interestingly, this scenario allows us to reconcile the simultaneous presence of infalling gas with roughly solar metallicity (Humphrey et al. 2007a, 2008b), which would correspond to the Rayleigh-Taylor 'fingers' left behind after the blow-out of the initial shell, and an expanding shell of low-metallicity gas at larger radii (e.g. Binette et al. 2000), corresponding to the second shell of swept-up matter.  Naively, one may otherwise have expected gas falling into the galaxy to have substantially lower metallicity than gas flowing out.  

We must also consider whether it is energeticaly feasible for the expansion of the absorbing shell to be starburst-driven.  Averaged across the HI, CIV, NV and SiIV lines (Table \ref{columns}), the current outflow velocity is $\sim$140 km s$^{-1}$.  Assuming this is a spherical shell of mass $\ge$1.6$\times$10$^{11} M_{\odot}$ ($\S$\ref{mass}), its kinetic energy would then be $\ge$3.1$\times$10$^{58}$ erg.  This serves as a lower limit on the mechanical energy deposited into the interstellar medium by supernovae.  Using Fig. 116 of Leitherer et al. (1999), we find that a modest star formation rate of $\sim$20 $M_{\odot}$ yr$^{-1}$, over a starburst lifetime of 100 Myr, would be sufficient to produce a mechanical energy luminosity of $\sim$3$\times$10$^{58}$ erg.  The total mass of stars formed in this starburst would be $\sim$2$\times$10$^9$ $M_{\odot}$, consistent with the stellar mass required to explain the alpha-element content of the shell ($\S$\ref{mass}).

Powerful radio sources can also provide a source of mechanical energy to potentially drive an outflow (see e.g. Nesvadba et al. 2006; Humphrey et al. 2010).  Could the radio source have produced this outflow?  The observed radius of the radio source is $\sim$3\arcsec (Carilli et al. 1997).  The fact that the radio source shows a relatively small arm-length asymmetry, shows a steep-spectrum and is lobe-dominated all imply that forshortening effects on its observed size can be neglected, so that its observed radius then corresponds to $\sim$25 kpc, which is substantially smaller than the r$\ge$40 kpc of the absorber.  While the ionizing radiation field of a quasar can also produce outflows (e.g. Humphrey et al. 2010), the kinetic energy of the shell around TXS 1436+157 is substantially higher than even the most luminous quasars can reasonably inject (e.g. Feruglio et al. 2010).  Thus, the expanding shell is unlikely to be driven by either the radio jets or the radiation field of the quasar.  

While the results presented in this paper directly concern only a single galaxy, they have wider implications for radio-loud active galaxies.  Most radio-loud active galaxies at high-z have 10-100 kpc-scale Ly$\alpha$ nebulae (e.g. Heckman et al. 1991; van Ojik et al. 1997) which are infalling where not perturbed by the radio jets (Humphrey et al. 2007).  A substantial fraction of these galaxies also show strong HI absorption features against the narrow Ly$\alpha$ emission line, which are usually blueshifted relative to the emission (van Ojik et al. 1997), and whose spatial extension indicates a location outside of the extended Ly$\alpha$ nebula.  Clearly, then, TXS 1436+157 is not unique in this regard, and we argue that the scenario we have proposed above should also be applicable to the wider population of high-z, radio-loud active galaxies.  

Finally, it is tempting to speculate that the shell may eventually escape from TXS 1436+157, and fragment into smaller self-gravitating baryonic entities, which at a later cosmic epoch may evolve into low-metallicity late-type galaxies.  These galaxies would be chemically pre-enriched to a level $12+log[O/H]\sim$7, in agreement with the low-metallicity theshold empirically found via observations of the most metal-poor star-forming galaxies currently known in the local Universe.  Such galaxies would presumably be almost free from dark matter, somewhat analogous to high-metallicity 'tidal dwarf galaxies' ejected from galaxy mergers (see the review by Duc 2010).  

\section{Conclusions}
Using deep long-slit optical spectroscopy from the 10.4 m GTC telescope, we have investigated the properties of narrow line emitting and absorbing gas associated with the z=2.54 radio-loud quasar TXS 1436+157.   Our main conclusions are as follows.  \\
\\
\noindent (i) The large Ly$\alpha$ nebula around the quasar appears to be infalling onto the host galaxy at a rate of $\sim$10-100 $M_{\odot}$ yr$^{-1}$.  It shows anomalously strong Ly$\alpha$ emission relative to HeII $\lambda$1640, due to enhancement of Ly$\alpha$ through resonance flourescence or through photoionization by stars (in addition to by the active nucleus).  \\
\\
\noindent (ii) The absorber lies outside of the Ly$\alpha$ nebula, and is likely to be a giant expanding shell enclosing the Ly$\alpha$ nebula and the host galaxy.  It has a hydrogen mass of $\ga 1.6\times10^{11}$ M$_{\odot}$, a gas density of $\le$18 cm$^{-3}$, a geometrical thickness of $\ge$18 pc, and a covering factor close to unity.  \\
\\
\noindent (iii) The absorber is detected in CIV, NV and SiIV lines, with low-ionization metal lines conspicuously absent, and is photoionized by the quasar.  Using photoionization models we conclude that the metallicity of the absorbing gas is $12+log[O/H]\le$7.3 -- much lower than that of the infalling Ly$\alpha$ nebula.  However, the detection of the metal lines shows that this is not pristine material.  \\
\\
\noindent (iv) To explain the properties of the Ly$\alpha$ nebula and the absorbing shell, we have proposed scenario in which a starburst-driven superbubble sweeps up material from the ISM; the bubble ruptures due to Rayleigh-Taylor instability, leaving behind filaments of cold, metal-enriched gas which fall back into the galaxy to fuel/trigger the quasar, while the hot supernova ejecta flows out into the low density halo and produces a new, larger shell of swept-up low metallicity gas.  \\
\\
\noindent (v) We argue that when illuminated by a sufficiently luminous quasar, gaseous shells such as this should be detectable in emission lines.

\section*{Acknowledgments}
Based on observations made during Mexican time with the Gran Telescopio Canarias at the Spanish Observatorio del Roque de los Muchachos, La Palma, Spain.  AH acknowledges a Marie Curie Fellowship cofunded by the 7$^{th}$ Research Framework Programme and the Portuguese Funda\c{c}\~ao para a Ci{\^e}ncia e a Tecnologia, and a \emph{CONACyT} post-doctoral research fellowship.  AH also acknowledges useful discussions with Patricio Lagos.  LB acknowledges support from \emph{CONACyT} grant CB-128556.  We also thank the anonymous referee for suggestions that improved this work.

\section*{References}

Allende Prieto C., Lambert D.~L., Asplund M., 2001, ApJ, 556, L63\\
\\
Anders E., Grevesse N., 1989, GeCoA, 53, 197\\
\\
Arshakian T.~G., Longair M.~S., 2000, MNRAS, 311, 846 \\
\\
Baker J.~C., Hunstead R.~W., Athreya R.~M., Barthel P.~D., de Silva E., 
Lehnert M.~D., Saunders R.~D.~E., 2002, ApJ, 568, 592\\
\\
Best P.~N., R{\"o}ttgering H.~J.~A., Longair M.~S., 2000, MNRAS, 311, 23 \\
\\
Best P.~N., Bailer D.~M., Longair M.~S., Riley J.~M., 1995, MNRAS, 275, 
1171 \\
\\
Binette L., Wilman R.~J., Villar-Mart{\'{\i}}n M., Fosbury R.~A.~E., Jarvis M.~J., R{\"o}ttgering H.~J.~A., 2006, A\&A, 459, 31\\
\\
Binette L., Kurk J.~D., Villar-Mart{\'{\i}}n M., R{\"o}ttgering H.~J.~A., 2000, A\&A, 356, 23\\
\\
Binette L., Dopita M.~A., Tuohy I.~R., 1985, ApJ, 297, 476\\
\\
Bridge C.~R., et al., 2012, arXiv, arXiv:1205.4030\\
\\
Carilli C.~L., R{\"o}ttgering H.~J.~A., van 
Ojik R., Miley G.~K., van Breugel W.~J.~M., 1997, ApJS, 109, 1\\
\\
Duc P.~A., 2010, in ``Dwarf Galaxies: Keys to Galaxy Formation and Evolution'', proceedings of JENAM, Lisbon, 2010, P. Papaderos, S. Recchi \& G. Hensler (eds.), Springer Verlag, p. 305  \\
\\
Ferruit P., Binette L., Sutherland R.~S., Pecontal E., 1997, A\&A, 322, 73\\
\\
Feruglio C., Maiolino R., Piconcelli E., Menci N., Aussel H., Lamastra A., Fiore F., 2010, A\&A, 518, L155 \\
\\
Fosbury R.~A.~E., et al., 2003, ApJ, 596, 
797\\
\\
Haiman Z., Rees M.~J., Loeb A., 1996, ApJ, 467, 522 \\
\\
Hayashi M., Kodama T., Tadaki K.-i., 
Koyama Y., Tanaka I., 2012, arXiv, arXiv:1207.2614 \\
\\
Hayes M., Scarlata C., Siana B., 2011, Natur, 476, 304 \\
\\
Heckman T.~M., Miley G.~K., Lehnert M.~D., 
van Breugel W., 1991, ApJ, 370, 78 \\
\\
Humphrey A., Villar-Mart{\'{\i}}n M., 
S{\'a}nchez S.~F., Mart{\'{\i}}nez-Sansigre A., Delgado R.~G., P{\'e}rez 
E., Tadhunter C., P{\'e}rez-Torres M.~A., 2010, MNRAS, 408, L1 \\
\\
Humphrey A., et al., 2008a, MNRAS, 390, 
1505 \\
\\
Humphrey A., Villar-Mart{\'{\i}}n M., 
Vernet J., Fosbury R., di Serego Alighieri S., Binette L., 2008b, MNRAS, 
383, 11 \\
\\
Humphrey A., Villar-Mart{\'{\i}}n M., 
Fosbury R., Binette L., Vernet J., De Breuck C., di Serego Alighieri S., 
2007, MNRAS, 375, 705 \\
\\
Humphrey A., Villar-Mart{\'{\i}}n M., 
Vernet J., Fosbury R., di Serego Alighieri S., Binette L., 2008b, MNRAS, 
383, 11\\
\\
Humphrey A., Villar-Mart{\'{\i}}n M., 
Fosbury R., Vernet J., di Serego Alighieri S., 2006, MNRAS, 369, 1103 \\
\\
Izotov Y.~I., et al., 2009, A\&A, 503, 61 \\
\\
Izotov Y.~I., Schaerer D., Blecha A., Royer F., Guseva N.~G., North P., 2006, A\&A, 459, 71 \\
\\
Izotov Y.~I., Chaffee F.~H., Foltz C.~B., Green R.~F., Guseva N.~G., Thuan 
T.~X., 1999, ApJ, 527, 757 
\\
Jarvis M.~J., Wilman R.~J., R{\"o}ttgering H.~J.~A., Binette L., 2003, 
MNRAS, 338, 263\\
\\
Krause M., 2002, A\&A, 386, L1 \\
\\
Leitherer C., et al., 1999, ApJS, 123, 3 \\
\\
McCarthy P.~J., Spinrad H., Djorgovski S., Strauss M.~A., van Breugel W., Liebert J., 1987, ApJ, 319, L39 \\
\\
McCarthy P.~J., 1993, ARA\&A, 31, 639\\
\\
Moll{\'a} M., Terlevich R., 2012, MNRAS, 425, 1696 \\
\\
Mori M., Umemura M., 2006, Natur, 440, 644\\
\\
Nesvadba N.~P.~H., Lehnert M.~D., 
Eisenhauer F., Gilbert A., Tecza M., Abuter R., 2006, ApJ, 650, 693 \\
\\
Papaderos P., Izotov Y.~I., Guseva N.~G., Thuan T.~X., Fricke K.~J., 2006, A\&A, 454, 119 \\
\\
Pettini M., Rix S.~A., Steidel C.~C., 
Adelberger K.~L., Hunt M.~P., Shapley A.~E., 2002, ApJ, 569, 742\\
\\
Reuland M., et al., 2003, ApJ, 592, 755\\
\\
Robinson A., Binette L., Fosbury R.~A.~E., 
Tadhunter C.~N., 1987, MNRAS, 227, 97\\
\\
R\"ottgering, H. J. A. 1993, Ph.D. thesis, Leiden Univ.\\
\\
R{\"o}ttgering H.~J.~A., van Ojik R., Miley G.~K., Chambers K.~C., van Breugel W.~J.~M., de Koff S., 1997, A\&A, 326, 505\\
\\
Rottgering H.~J.~A., Hunstead R.~W., Miley 
G.~K., van Ojik R., Wieringa M.~H., 1995a, MNRAS, 277, 389\\
\\
R{\"o}ttgering H.~J.~A., Miley G.~K., Chambers K.~C., Macchetto F., 1995b, A\&AS, 114, 51 \\
\\
Spitzer L., 1978, Physical processes in the interstellar medium, Wiley-Interscience, New York\\
\\
Steidel C.~C., Bogosavljevi{\'c} M., 
Shapley A.~E., Kollmeier J.~A., Reddy N.~A., Erb D.~K., Pettini M., 2011, 
ApJ, 736, 160\\
\\
Steidel C.~C., Adelberger K.~L., Shapley 
A.~E., Pettini M., Dickinson M., Giavalisco M., 2000, ApJ, 532, 170\\
\\
Tenorio-Tagle G., Silich S.~A., Kunth D., 
Terlevich E., Terlevich R., 1999, MNRAS, 309, 332 \\
\\
Thuan T.~X., Lecavelier des Etangs A., Izotov Y.~I., 2005, ApJ, 621, 269 \\
\\
van Ojik R., Roettgering H.~J.~A., Miley G.~K., Hunstead R.~W., 1997, A\&A, 317, 358\\
\\
Vernet J., Fosbury R.~A.~E., Villar-Mart{\'{\i}}n M., Cohen M.~H., Cimatti A., di Serego Alighieri S., Goodrich R.~W., 2001, A\&A, 366, 7 \\
\\
Villar-Mart{\'{\i}}n M., S{\'a}nchez 
S.~F., Humphrey A., Dijkstra M., di Serego Alighieri S., De Breuck C., 
Gonz{\'a}lez Delgado R., 2007a, MNRAS, 378, 416 \\
\\
Villar-Mart{\'{\i}}n M., Humphrey A., De 
Breuck C., Fosbury R., Binette L., Vernet J., 2007b, MNRAS, 375, 1299 \\
\\
Villar-Mart{\'{\i}}n M., Vernet J., di 
Serego Alighieri S., Fosbury R., Humphrey A., Pentericci L., 2003, MNRAS, 
346, 273 \\
\\
Villar-Mart{\'{\i}}n M., Vernet J., di 
Serego Alighieri S., Fosbury R., Pentericci L., Cohen M., Goodrich R., 
Humphrey A., 2002, MNRAS, 336, 436 \\
\\
Webb T.~M.~A., Yamada T., Huang J.-S., Ashby M.~L.~N., Matsuda Y., Egami 
E., Gonzalez M., Hayashimo T., 2009, ApJ, 692, 1561\\
\\
Weidinger M., M{\o}ller P., Fynbo J.~P.~U., 2004, Natur, 430, 999 \\
\\
Wilman R.~J., Gerssen J., Bower R.~G., Morris S.~L., Bacon R., de Zeeuw 
P.~T., Davies R.~L., 2005, Natur, 436, 227\\
\\
Wilman R.~J., Jarvis M.~J., R{\"o}ttgering H.~J.~A., Binette L., 2004, 
MNRAS, 351, 1109

\end{document}